\documentclass[format=acmsmall, anonymous=false, natbib=true]{acmart}
\renewcommand\footnotetextcopyrightpermission[1]{}
\usepackage{booktabs} 
\usepackage[utf8]{inputenc}
\usepackage[T1]{fontenc}

\usepackage{hyperref}
\usepackage{cleveref}

\usepackage{graphicx}
\usepackage{subcaption}
\usepackage{xcolor}
\usepackage[acronym]{glossaries}
\usepackage{multirow}
\usepackage{array}
\usepackage{float}
\usepackage{tikz}
\usetikzlibrary{arrows,positioning}

\graphicspath{{figures/}}
\makeatletter
\def\input@path{{figures/}}

\usepackage[para]{threeparttable}
\usepackage{tablefootnote}

\newacronym{ar}{AR}{affect recognition}
\newacronym{har}{HAR}{human activity recognition}
\newacronym{ecg}{ECG}{electrocardiogram}
\newacronym{skt}{TEMP}{skin temperature}
\newacronym{eda}{EDA}{electrodermal activity}
\newacronym{gsr}{GSR}{galvanic skin response}
\newacronym{emg}{EMG}{electromyogram}
\newacronym{resp}{RESP}{respiration}
\newacronym{acc}{ACC}{3-axis accelerometer}
\newacronym{bp}{BP}{blood pressure}
\newacronym{bvp}{BVP}{blood volume pulse}
\newacronym{eeg}{EEG}{electroencephalogram}
\newacronym{ppg}{PPG}{photoplethysmography}
\newacronym{eog}{EOG}{electrooculography}
\newacronym{ans}{ANS}{autonomous nervous system}
\newacronym{sns}{SNS}{sympathetic nervous system}
\newacronym{pns}{PNS}{parasympathetic nervous system}
\newacronym{scl}{SCL}{skin conductance level}
\newacronym{scr}{SCR}{skin conductance response}
\newacronym{nscr}{nSCR}{number of skin conductance responses}
\newacronym{ema}{EMA}{ecological momentary assessment}
\newacronym{ema_pl}{EMAs}{ecological momentary assessments}
\newacronym{pd}{PD}{pupil diameter}
\newacronym{rip}{RIP}{respiratory inductive plethysmograph}

\newacronym{iaps}{IAPS}{International Affective Picture System}
\newacronym{tsst}{TSST}{Trier Social Stress Test}
\newacronym{mist}{MIST}{Montreal Imaging Stress Task}

\newacronym{pss}{PSS}{Perceived Stress Scale}
\newacronym{panas}{PANAS}{Positive and Negative Affect Schedule}
\newacronym{bfi}{BFI}{Big Five Inventory}
\newacronym{bfms}{BFMS}{Big Five Marker Scale}
\newacronym{psqi}{PSQI}{Pittsburgh Sleep Quality Index}
\newacronym{phq}{PHQ-9}{Patient Health Questionnaire}
\newacronym{stai}{STAI}{Strait-Trait Anxiety Inventory}
\newacronym{sri}{SRI}{Stress Response Inventory}
\newacronym{pam}{PAM}{Photo Affect Meter}
\newacronym{sam}{SAM}{Self-Assessment Manikins}
\newacronym{pca}{PCA}{Principal Component Analysis}
\newacronym{loo}{LOO}{Leave-One-Out}
\newacronym{loso}{LOSO}{Leave-One-Subject-Out}
\newacronym{lda}{LDA}{Linear Discriminant Analysis}
\newacronym{qda}{QDA}{Quadratic Discriminant Analysis}
\newacronym{knn}{kNN}{k-Nearest Neighbour}
\newacronym{svm}{SVM}{support vector machine}
\newacronym{rf}{RF}{Random Forest}
\newacronym{cv}{CV}{Cross-Validation}
\newacronym{loto}{LOTO}{Leave-One-Trial-Out}
\newacronym{nn}{NN}{neural networks}
\newacronym{cnn}{CNN}{convolutional neural network}
\newacronym{lstm}{LSTM}{long short-term memory}
\newcommand{\tabitem}{~~\llap{\textbullet}~~}

\newacronym{hr}{HR}{heart rate}
\newacronym{hrv}{HRV}{heart rate variability}
\newacronym{psd}{PSD}{power spectral density}
\newacronym{ibi}{IBI}{inter beat interval}
\newacronym{rsa}{RSA}{respiratory sinus arrhythmia}

\begin{document}
\title{Wearable affect and stress recognition: A review}

\author{Philip Schmidt}
\email{philip.schmidt@de.bosch.com}

\author{Attila Reiss}
\email{attila.reiss@de.bosch.com}

\author{Robert Duerichen}
\email{robert.duerichen@de.bosch.com}

\affiliation{%
	\institution{Robert Bosch GmbH}
	\streetaddress{Robert-Bosch-Campus 1}
	\city{Renningen}
	\postcode{71272}
	\country{Germany}
}

\author{Kristof Van Laerhoven}
\affiliation{%
	\institution{University Siegen}
	\streetaddress{H\"olderlinstr. 3}
	\city{Siegen}
	\country{Germany}}
\email{kvl@eti.uni-siegen.de}
\renewcommand{\shortauthors}{P. Schmidt et al.}

\settopmatter{printacmref=false}
\setcopyright{none}

%
%

\ccsdesc[500]{Human-centered computing~Ubiquitous and mobile computing}

%
%
\keywords{Affective computing, affect recognition, wearables, user studies, survey}

\begin{abstract}
	Affect recognition aims to detect a person's affective state based on observables, with the goal to e.g. provide reasoning for decision making or support mental wellbeing. 
	Recently, besides approaches based on audio, visual or text information, solutions relying on wearable sensors as observables (recording mainly physiological and inertial parameters) have received increasing attention.
	Wearable systems offer an ideal platform for long-term affect recognition applications due to their rich functionality and form factor. 
	However, existing literature lacks a comprehensive overview of state-of-the-art research in wearable-based affect recognition.
	Therefore, the aim of this paper is to provide a broad overview and in-depth understanding of the theoretical background, methods, and best practices of wearable affect and stress recognition.
	We summarise psychological models, and detail affect-related physiological changes and their measurement with wearables.
	We outline lab protocols eliciting affective states, and provide guidelines for ground truth generation in field studies.
	We also describe the standard data processing chain, and review common approaches to preprocessing, feature extraction, and classification.
	By providing a comprehensive summary of the state-of-the-art and guidelines to various aspects, we would like to enable other researchers in the field of affect recognition to conduct and evaluate user studies and develop wearable systems.
\end{abstract}

\maketitle
\thispagestyle{empty}

\section{Introduction}
\label{sec:intro}
\Glsdesc{ar} aspires to detect the affective state of a person based on observables.
Hence, from a theoretical point of view, \glsdesc{ar} can be seen as a signal and pattern recognition problem\cite{2015_Dmello}.
From a practical standpoint, \glsdesc{ar} is an essential building block of affective computing (e.g. affect-aware interfaces).
As a result, affect recognition is a highly interdisciplinary research field with links to signal processing, machine learning, psychology, and neuroscience.

Experiments of \citet{1981_Bower} indicate that decision making and memorisation of a person are strongly influenced by their affective states.
Therefore, a holistic user model requires the affective state as an integral part.
Such a model could not only provide reasoning for the user's actions, but also be of great value to the user by providing insights into his/her affective states.
Correlations between certain affective states (e.g. joy) and places (e.g. cinema) or persons (e.g. friends) could be helpful for users when planning their leisure activities. 
From a healthcare point of view, stress is a particularly interesting affective state.
This is due to the severe side effects of long-term stress, which range from headaches and troubled sleeping to an increased risk of cardiovascular diseases \cite{1993_McEwen, 1992_Chrousos, 1998_Rosmond}.
According to the British Health and Safety Executive (HSE), stress accounted for 37\% of all work-related ill health cases in 2015/16\cite{stress}.
As a result, a frequently pursued task in \glsdesc{ar} is to build automated stress detection systems.

In the affect recognition literature, numerous approaches based on audio-visual data\cite{2017_Tzirakis, 2017_Mirsamadi}, contextual cues\cite{2014_Wang}, text\cite{2014_Gangemi}, and physiology\cite{2001_Picard,2017a_Gjoreski} have been presented.
In this review, we focus on approaches utilising wearable sensors (recording mainly physiological and inertial parameters).
The reasons for this focus are twofold:
First, due to their rich functionality and form factor, wearables like smart phones/watches are popular among users.
A clear goal of \glsdesc{ar} systems is to be applicable in everyday life.
Due to their computational power and integrated sensors, wearable devices are ideal platforms for many applications.
Consumer wearables already offer activity recognition, and recently a first generation of affect (e.g. stress) recognition systems entered in this sector\cite{garmin}. 
Second, parameters observable with wearable sensors (such as changes related to the cardiac system or electrodermal activity) provide valuable insights related to the user's affective state.
Moreover, most related work relies on a multimodal setup.
\citet{2015_Dmello} pointed out that \glsdesc{ar} systems basing their decisions on multimodal data tend to be almost 10\% more accurate than their unimodal counterparts. 

\glsreset{ar}
The aim of this work is to provide a broad overview and in-depth understanding of the theoretical background, methods, and best practices of wearable affect and stress recognition.
By providing a comprehensive summary of the state-of-the-art, we would like to enable other researchers in the field of affect recognition to conduct and evaluate user studies and develop wearable systems.
Since the focus is on wearable solutions, approaches and studies relying mainly on audio, video, or text information are not subject of this review.
Although \glsdesc{ar} systems based on audio-visual data are very powerful and incorporated in products (e.g. Affectiva\cite{affectiva}), we exclude these modalities due to their limitations regarding mobile systems for everyday life and their intrusive nature.
We refer readers with an interest in \glsdesc{ar} or sentiment analysis methods based on audio or visual data to \citet{2017_Poria}.
Moreover, work relying solely or mainly on smart phone data is excluded as well, since we focus on approaches relying on the observation of physiological changes of the user.
Details concerning affect recognition based on smart phone usage can be found in \citet{2012_Miller}.
Finally, we exclude the vast work done in the field of \gls{eeg} based affect recognition due to the practical limitations of applying \gls{eeg} in real-life scenarios.
\Gls{eeg}-based affect recognition is reviewed e.g. by \citet{2013_Kim}.

The rest of this review is organised as follows.
In \Cref{se:stress_emo}, psychological models of affect are presented.
Then, the influence of different affective states on the human physiology, and the sensors commonly used to measure physiological states and changes are detailed in \Cref{sec:physio_affect}.
Next, guidelines for laboratory and field studies are presented in \Cref{sec:data_gt}.
For this purpose, we outline standardised lab protocols eliciting affective states and address the issue of ground truth generation in the field.
Furthermore, \Cref{sec:data_set} details publicly available data sets, containing wearable sensor data recorded in response to an affective stimulus.
\Cref{sec:feat_proc} provides a detailed description of the standard data processing chain, which is frequently employed to associate raw wearable data with different affective states.
We overview common approaches related to the steps preprocessing, feature extraction, and classification.
Finally, this work is concluded in \Cref{sec:discussion} by summarising the main findings and outlining future challenges in wearable-based \glsdesc{ar}.

\section{Psychological background}
\label{se:stress_emo}
\glsreset{ar}
This section provides a brief psychological background and overview of psychological models used in affect and stress recognition.
First, \Cref{sub:basdef} defines the terms affect, emotion, and mood.
Then, \Cref{sec:emo_mod} provides a summary of categorical and dimensional models of emotions.
Finally, in \Cref{sec:stress} different stress concepts will be presented.

\subsection{Affect, emotion, mood - basic definitions}
\label{sub:basdef}
Despite a growing body of research, it is still difficult to define the terms affect, emotion, and mood in a precise way.
Below we provide working definitions for these terms, and highlight the differences between them. 
\citet{2003_Russell} defines \textbf{affect} to be a neurophysiological state. 
This neurophysiological state is consciously accessible as simple raw (nonreflective) primitive feeling\cite{2017_Liu}.
Affect is not directed at a specific event or object and lasts only for a very short time.
In contrast, \textbf{emotions} are intense and directed feelings, which have a short duration.
Emotions are an indicator of affect, and arise from a cognitive process evaluating a stimulus (e.g. specific objects, an affect, a thought).
Therefore, emotions are directed at a stimulus. 
To illustrate these aspects, \citet{2017_Liu} uses the example of watching a scary movie:
If you are affected, the movie elicits the feeling of being scared.
The mind processes this feeling (\textit{scared}), adds an evaluation (e.g. 'this is really spooky'), and expresses it to you and your surroundings as an emotion (\textit{fear}) by e.g. crying\cite{2017_Liu}.
In the affect recognition literature, the terms mood and emotion are often used interchangeably.
However, in contrast to emotion (and affect),  \textbf{mood} is commonly defined to be less intense, more diffuse, and lasts for a longer time period.
This difference between mood and emotion is best illustrated by considering the following example: One can get angry very quickly, but it is hard to stay angry for a longer time period.
However, the emotion \textit{anger} might lead to an \textit{irritable} mood, which can last for a long time\cite{2017_Liu}.
In the remainder of this paper we will use the term \textbf{affective state} to describe the internal state of a person, which can be referred to as emotion, mood, and/or affect.


\subsection{Emotion models}
\label{sec:emo_mod}
\glsreset{ar}
In this section the emotional models frequently employed in the affect recognition literature are detailed. 
The models used can be grouped into two different families:
\begin{enumerate}
	\item \textbf{Categorical models:} Different emotions are represented best by discrete categories. 
	\item \textbf{Dimensional models:} Emotions can be mapped into a multidimensional space.
\end{enumerate}
\textbf{Categorical models} date back to ancient Greek and Roman philosophers\cite{2017_Poria}.
Cicero, for instance, distinguished four basic categories of emotions, namely \textit{fear, pain, lust}, and \textit{pleasure}\cite{2002_Cicero}. 
\citet{1872_Darwin} also conducted studies on emotions, and came to the conclusion that emotions have an evolutionary history and hence are shared across cultures. 
Similar to \citeauthor{1872_Darwin}, \citet{1992_Ekman} argues that basic emotions are shared across cultures and appear to be universally recognised.
Following \citeauthor{1978_Ekman}, six basic emotions can be distinguished: \textit{joy, sadness, anger, fear, disgust,} and \textit{surprise} \cite{1978_Ekman, 1976_Ekman, 1992_Ekman}. 
These basic emotions are discrete and have distinct physiological patterns (e.g. facial muscle movement).
According to \citet{1992_Ekman}, nine characteristics can be used in total to distinguish basic emotions from one another, and from other affective phenomena.
Being able to express basic emotions can be attributed with a number of (evolutionary evolved) physiological and communicative functions. 
Consider for instance the facial expression of \textit{disgust}:  
On a physiological level the wrinkled nose, often expressing disgust, limits the intake of malodorous air/particles.
Being able to recognise this facial expression has on the communicative level the potential to warn others of dangerous (e.g. poisonous) food, which can make the difference between life and death.

\begin{figure}[t]
\begin{minipage}{.47\textwidth}
	\centering
	\includegraphics[width=0.7\textwidth]{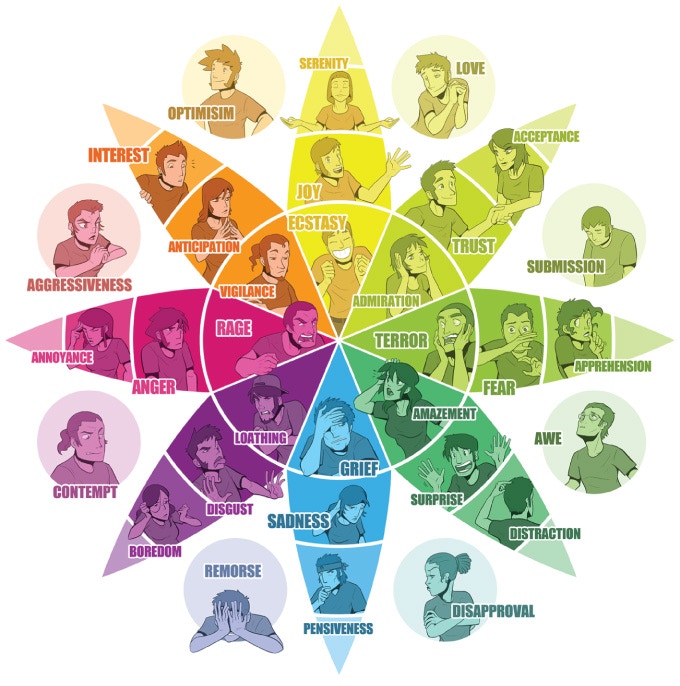}
	\caption{Schematic representation of \newline the 'wheel of emotions'\cite{kick_wheel}.}
	\label{fig:wheel}
\end{minipage}%
\begin{minipage}{.47\textwidth}
	\centering
	\includegraphics[width=0.9\textwidth]{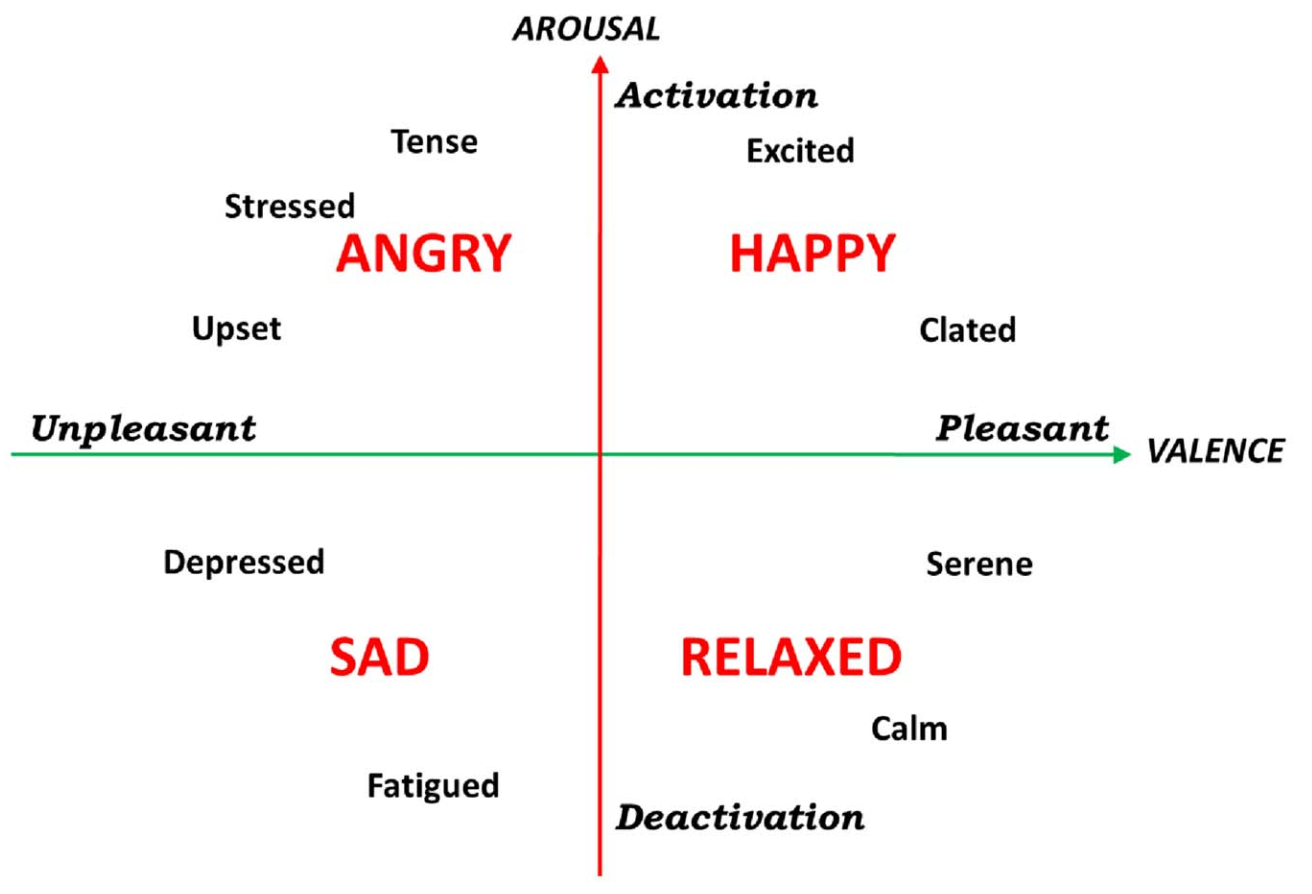}
	\vspace{1.5em}
	\caption{Schematic representation of the circumplex model. Adapted from \citet{2014_Valenza}.}
	\label{fig:val_ar}
\end{minipage}
\vspace{-1em}
\end{figure}

\begin{figure}[t]
	\begin{minipage}{.47\textwidth}
		\centering
		\includegraphics[width=\textwidth]{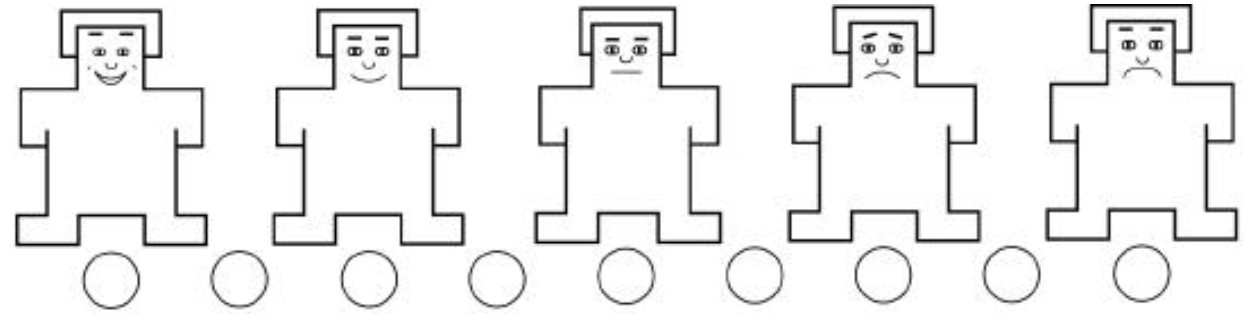}
	\end{minipage}%
	\begin{minipage}{.47\textwidth}
		\vrule
		\centering
		\vspace{0.35em}
		\includegraphics[width=0.97\textwidth]{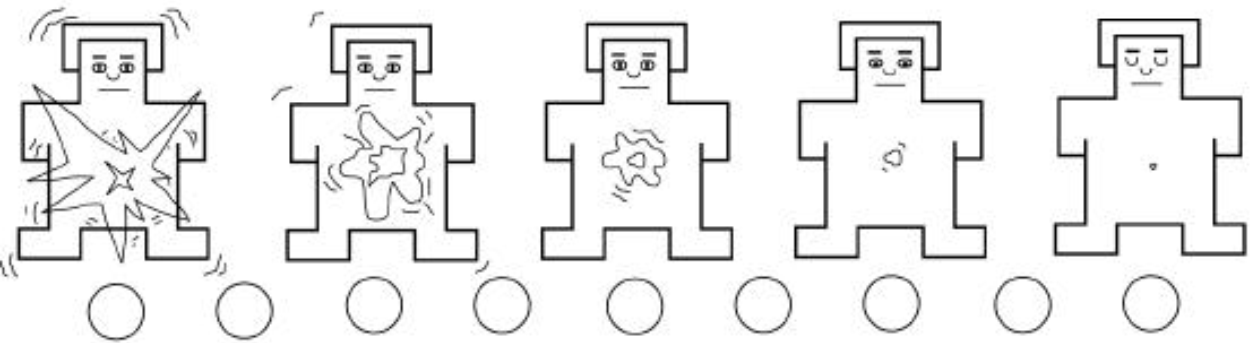}
	\end{minipage}
	\caption{\glsdesc{sam} (Left: valence, Right: arousal) adapted from \citet{1995_Morris}.}
	\label{fig:sam}
	\vspace{-1em}
\end{figure}

In 1980, \citet{1980_Plutchik} developed a taxonomy to classify discrete emotions.
The so-called 'wheel of emotions' comprises of eight primary emotions: \textit{grief, amazement, terror, admiration, ecstasy, vigilance, rage,} and \textit{loathing}. 
Following \citet{1980_Plutchik}, the primary emotions mix, and hence give rise to new, more complex emotions.
In addition, emotions can be  expressed at different intensity levels.
\autoref{fig:wheel} depicts a schematic image of the 'wheel of emotions'.
In the domain of wearable affect recognition, categorical models are for instance used by \citet{2016_Zenonos}:
They distinguish eight different emotions and moods (\textit{excited, happy, calm, tired, bored, sad, stressed} and \textit{angry}).\\
\textbf{Dimensional models} allow the mapping of emotions into a multidimensional space.
The first dimensional approach dates back to \citet{1863_Wundt}, who describes momentary emotions as a single point in a three-dimensional space\cite{2008_Becker}.
This 'emotional space' is spanned by the pleasure-displeasure, excitement-inhibition, and tension-relaxation axes.
At the end of the 1970s, \citet{1979_Russell} postulated a two-dimensional model, namely the circumplex model. 
In this model, affective states are represented as discrete points in a two-dimensional space, spanned by the axes valence and arousal. 
The valence axis indicates how positive or negative the currently experienced state is.
On the arousal axis, the state is rated in terms of the activation level (e.g. how energised or enervated one feels).
\autoref{fig:val_ar} displays Russell's circumplex model schematically.
The four quadrants of the circumplex model can be attributed with \textit{happy, angry, sad,} and \textit{relaxed}.
By adding further orthogonal axes (e.g. dominance), the circumplex model is easily extended\cite{2012_Koelstra, 2015_Abadi}.
In affect recognition, the circumplex model is the most frequently employed model to capture the affective state of a subject\cite{2008_AKim, 2014_Valenza, 2015_Abadi}. 
One reason for the popularity of the circumplex model is that it can be easily assessed using the so-called \gls{sam}\cite{1995_Morris}.
These Manikins offer an easy graphical way to report current affective states (see \autoref{fig:sam}).
In addition, due to their simple graphical representation, \gls{sam} are easily understood across cultures.
Another possible reason for the popularity of dimensional models in \glsdesc{ar} might arise from a machine learning point of view:
The (at least two) independent axes of the circumplex model offer an interesting set of different classification problems.
For instance, the values on the valence and arousal axes can be cast into multiclass classification problems (e.g. low/medium/high arousal or valence).
The four quadrants in the space spanned by the valence and arousal axes can also be used to define a classification task\cite{2008_AKim}.

\subsection{Stress models}
\label{sec:stress}
\glsreset{ar}
In everyday life \textit{stress} or \textit{being stressed} are terms used to describe the feeling of being under pressure.
Stress is commonly elicited by an external and/or internal stimulus called stressor.
However, from a scientific point of view, stress is primarily a physiological response.
A pure physiological view on stress dates back to the beginning of the 20th century when \citet{1929_Cannon} coined the term homeostasis.
According to \citeauthor{1929_Cannon}, homeostasis defines the strive of an organism to keep its' physiological parameters (e.g. blood glucose) within an acceptable range.
Following \citeauthor{1929_Cannon}, both physiological and psychological stimuli can pose threats to homeostasis.
In order to maintain homeostasis, even under extreme conditions, feedback loops are triggered when the current (physiological) state lies outside of the acceptable parameter boundaries.

In the 1970s, \citet{1974_Selye} defined stress to be (or result in) a 'nonspecific response of the body to any demand upon it'.
Following this definition, 'nonspecific' refers to a shared set of responses triggered regardless of the nature of the stressor (e.g. physical or psychological). 
Recent stress models, see e.g. \citet{1993_McEwen}, incorporate multiple effectors and advocate that the stress response is to some degree specific.
The stress response is mainly influenced by two aspects:
first the stressor itself and second the organism's perceived ability to cope with the threat\cite{2007_Goldstein}. 
Depending on the coping ability of the organism and perceived chances for success, \textbf{eustress} (positive outcome) and \textbf{distress} (negative outcome) are distinguished\cite{2012_Lu}.
Eustress can have a positive (e.g. motivating) effect on a person, while distress can be hindering (feeling e.g. worried or anxious).
Consider the following example: Assume a person has to take an exam, then the exam can be interpreted as an external stressor.
The body reacts to this stressor by providing more energy (physiological stress response).
If the person feels well prepared for the exam and is looking forward to the challenge ahead, this can be interpreted as eustress.
In contrast, if the person is not well prepared and feels like failing the exam, this can be seen as distress.
Considering wearable stress recognition, distinguishing between eustress and distress is a largely unsolved problem due to the lack of e.g. adequate physiological indicators.
Therefore, most work in this area defines either a binary stress recognition task (\textit{stress} vs. \textit{no stress})\cite{2017_Mozos, 2011_Plarre} or aims at distinguishing different levels of stress (e.g. \textit{no stress} - \textit{low stress} - \textit{high stress})\cite{2017a_Gjoreski}.


In \Cref{sec:emo_mod} different emotion models were summarised.
Although stress is not a basic emotion, a link between dimensional models and stress is readily established.
Following \citet{2010_Sanches}, a direct link between stress and arousal can be drawn.
\citet{2014_Valenza} maps stress into the high arousal/negative valence (quadrant II) of the circumplex model (see \autoref{fig:val_ar}).
Following \citet{1990_Thayer} and later \citet{2002_Schimmack}, the arousal dimension of the 'classical circumplex' model can be split into tense arousal (stressed-relaxed) and energetic arousal (sleepy-active).
According to \citet{2002_Schimmack}, this split is justified by the observation that only the energetic arousal component is influenced by the sleep-wake cycle.
Considering the wearable affect and stress recognition literature, a recent study conducted by \citet{2017_Mehrotra} uses this three-dimensional emotion model (valence, tense arousal, energetic arousal) to investigate correlation and causation between emotional states and mobile phone interaction.

\section{Affective states: physiological changes and their measurement}
\label{sec:physio_affect}
This section provides background on the physiological changes associated with affective states (\Cref{sub:af_foot}), and details sensor modalities frequently employed for wearable affect recognition to measure these changes (\Cref{sub:sensors}).
Furthermore, \Cref{sub:freq_em_sensors} summarises studies and related work with respect to the targeted affective states and included sensor modalities.

\subsection{Affective states and their physiological fingerprint}
\label{sub:af_foot}
\glsreset{ar}
There is clearly a link between affective states and physiological changes.
For example, if someone tells us a good joke we laugh or at least smile.
With this physiological response we express e.g. \textit{amusement}.
Negative emotional states might even have stronger physiological indicators.
For instance, when being \textit{afraid} or \textit{anxious} one might start sweating, get a dry mouth, or feel sick.
As mentioned in \Cref{sec:stress}, stress is characterised primarily as a physiological response to a stimulus.
The most severe physiological reaction to a stressor is the so called 'fight or flight' response\cite{1929_Cannon}.
During the 'fight or flight' response a mixture of hormones (e.g. cortisol and adrenaline) are released, leading to an increased breathing/heart rate and muscle tension.
These physiological changes prepare the organism for a physical reaction, which can make the difference between life and death. 
Hence, the induced physiological responses are quite distinct.

The above examples demonstrate the evident link between affective states and physiological responses.
The direction (causality) of this link is, however, still an open research question.
Common sense tells us that emotions cause bodily changes.
However, \citet{1884_James} postulated, at the end of the 19th century, that emotions arise from the changes in physiology.
The following example illustrates this theory: Imagine encountering a big black poisonous spider.
Probably your heart would start to race and you start to sweat. 
Following \citet{1884_James}, these physiological changes are not symptoms of \textit{fear}, but rather involuntary physiological responses (affects) that only become an emotion (e.g. \textit{fear}) by adding a cognitive evaluation.
This feedback theory is supported by an experiment conducted by \citet{1990_Levenson}.
The experiment found evidence that performing voluntary facial muscle movements exhibit similar changes in peripheral physiology as if the emotion is experienced.
For instance, when the subjects were asked to make an angry face the \glsdesc{hr} was found to increase.
The debate outlined above is, from a theoretical point of view, very interesting. 
However, the question if an emotion triggers a physiological response or vice versa is out of the scope of this review. 
For affect and stress recognition based on wearable sensors, the fact that affective states are accompanied by more or less pronounced physiological changes is essential.

\begin{table}[t]
	\centering
	\footnotesize
	\caption{Overview of the two branches of the \glsdesc{ans} and their major functions.}
	\vspace{-1em}
	\begin{tabular}{|l|l|}
		\hline
		\textbf{Sympathetic nervous system (SNS)} 						& \textbf{Parasympathetic nervous system (PNS)}\\ \hline
		\tabitem associated with 'fight or flight'						& \tabitem associated with 'rest and digest'\\
		\tabitem pupils dilate 											& \tabitem pupils constrict\\
		\tabitem decreased salivation and digestion						& \tabitem increased salivation and digestion\\
		\tabitem increased heart and respiration rate					&\tabitem decreased heart and respiration rate \\
		\tabitem increased electrodermal activity and muscle activity	&\\
		\tabitem adrenalin and glucose release					&\\
		\hline
	\end{tabular}
	\label{tab:sns_pns}
	\vspace{-2em}
\end{table}

Affective states occur spontaneously and are accompanied by certain physiological responses, which are hard if not impossible to control for humans. 
The \gls{ans} directs the unconscious actions of the organism.
Hence, the \gls{ans} plays a key role in directing the physiological response to an external (e.g. event) or internal (e.g. thought) affective stimulus.
The \gls{ans} has two major branches: the \gls{sns} and the \gls{pns}.
In \autoref{tab:sns_pns}, the key contributions of the \gls{sns} and \gls{pns} are displayed.
As the \gls{sns} is mainly associated with the 'fight-or-flight' response, an increased activity of the \gls{sns} indicates higher arousal states.
Concluding from \autoref{tab:sns_pns}, the \gls{sns} has the function to provide energy by increasing a number of physiological parameters (e.g. respiration rate, glucose level).
The \gls{pns}, in contrast, regulates the 'rest and digest' functions\cite{2007_McCorry}.

The interplay of \gls{sns} and \gls{pns} is best illustrated considering the cardiovascular system.
In reaction to a potential threat, the \gls{sns} increases the \gls{hr}.
Once the threat is over, the \gls{pns} reduces the \gls{hr}, bringing it back to a rest state\cite{2012_Choi}.
A common measure to quantify the interaction of \gls{sns} and \gls{pns} is the \gls{hrv}.
The \gls{hrv} is defined as the variation in the beat-to-beat intervals.
An increased/decreased \gls{hrv} indicates increased activity of the \gls{pns}/\gls{sns}, respectively.
As a result, the \gls{hrv} is a rather simple but efficient measure to quantify the contributions of the \gls{pns}/\gls{sns}, and e.g. to deduce the stress level.
Changes in the \gls{eda} are a simple but yet effective measure to assess the \gls{sns} activity.
This is due to the fact, that the \gls{eda} is solely stimulated by the \gls{sns}\cite{2012_Choi}.
Hence, \gls{eda} is particularly sensitive to high arousal states\cite{2000_Dawson} (e.g. \textit{fear, anger, stress}).
\gls{eda} has two main contributions, namely the \gls{scl} and the \gls{scr}.
The \gls{scl} (also known as tonic component) represents a lowly varying base line conductivity.
In contrast, the \gls{scr} (phasic component) refers to peaks in the \gls{eda} signal.
For most other vital parameters, the contribution of \gls{pns} and \gls{sns} is more interleaved.
Hence, their responses are less specific. 
Nevertheless, also considering respiration and muscle activity, certain patterns can be attributed to different affective states.
For instance, the respiration rate increases and becomes more irregular when a subject is more aroused\cite{2008_AKim}.

\begin{table}[t]
	\centering
	\footnotesize
	\caption{Four different affective states and their physiological response\cite{2010_Kreibig}. $\downarrow$ indicates a decrease, $\uparrow$ indicates an increase, $\uparrow \downarrow$ indicates both increase and decrease (depending on the study), $-$ indicates no change in the parameter under consideration.}
	\vspace{-1em}
	\begin{tabular}{|l|c|c||c|c|}
		\hline
		\multirow{3}{*}{} & \multirow{3}{*}{Anger} & \multirow{2}{*}{Sadness} & \multirow{3}{*}{Amusement} & \multirow{3}{*}{Happiness} \\ 
		&&\multirow{2}{*}{(non-crying)}&&\\
		&&&&\\ \hline
		\multicolumn{5}{|l|}{\textbf{Cardiovascular:}}\\ \hline
		\Glsdesc{hr} (HR) &$\uparrow$&$\downarrow$& $\uparrow \downarrow$ & $\uparrow$\\ \hline
		\Glsdesc{hrv} (HRV) &$\downarrow$&$\downarrow$& $\uparrow$ &$\downarrow$\\ \hline\hline
		\multicolumn{5}{|l|}{\textbf{Electrodermal:}}\\ \hline
		\Glsdesc{scl} (SCL) &$\uparrow$&$\downarrow$&$\uparrow$& $\uparrow -$\\ \hline
		Number of \glsdesc{scr}s (SCR) &$\uparrow$&$\downarrow$&$\uparrow$&$\uparrow$\\ \hline\hline
		\multicolumn{5}{|l|}{\textbf{Respiration:}}\\ \hline
		Respiration rate &$\uparrow$& $\uparrow$&$\uparrow$&$\uparrow$\\ \hline
	\end{tabular}
	\vspace{-1.5em}
	\label{tab:effect_of_emotions}
\end{table}
As outlined above, the \gls{sns} contributions to high arousal states are quite distinct.
In a recent meta analysis, \citet{2010_Kreibig} investigated the specificity of the \gls{ans} reaction considering certain affective states.
The findings suggest, that the reactions of the \gls{ans} are to some extent specific.
A subset of these findings, including two positive and two negative affective states, is presented in \autoref{tab:effect_of_emotions}.
Considering \textit{anger}, a majority of the analysed studies showed that it coincides with an increased \gls{hr}, \gls{scl}, number of \gls{scr}s, and a higher breathing rate.
Since \textit{anger} represents a high arousal state, governed by the \gls{sns}, these reactions were expected.
Non-crying \textit{sadness} was found to decrease \gls{hr}, \gls{scl} and number of \gls{scr}s, while increasing the respiration rate.
In \autoref{fig:val_ar}, \textit{sadness} is mapped into the third quadrant. 
Hence, the arousal level is expected to drop which is confirmed by \autoref{tab:effect_of_emotions}.
\textit{Amusement} and \textit{happiness} are both positive affective states with a similar arousal level.
Hence, it is not surprising that they have a similar physiological fingerprint.
For more information about the specificity of \gls{ans} response, we refer the reader to \citet{2010_Kreibig}.

The findings of \citet{2010_Kreibig} suggest that affective states have certain physiological fingerprints which are to some degree specific.
These findings are promising, as they indicate that distinguishing affective states based on physiological indicators is feasible.
However, the following aspects should be considered\cite{2009_Broek}:
\vspace{-0.2em}
\begin{enumerate}
	\item Physiological measures are \textit{indirect} measures of an affective state.
	\item Emotions are subjective, but physiological data are not.
	\item Although some physiological patterns are shared across subjects, individual responses to a stimulus can differ strongly.
	\item Multimodal affect detecting systems reach higher accuracies than unimodal systems\cite{2015_Dmello}.
	\item The physiological signal quality often suffers from noise, induced by motion artefacts and misplacement.
\end{enumerate}
\vspace{-0.5em}

\subsection{Sensors}
\label{sub:sensors}
\glsreset{ar} 
This section provides an overview of the sensor modalities frequently employed in the wearable affect and stress recognition literature.
The clear aim of these work is to find robust methods assessing the affective state of a user in everyday life.
Hence, a major goal is to use sensor setups which are minimally intrusive and pose only minor limitations to the mobility of the user.
As defined in \autoref{tab:sns_pns} and \autoref{tab:effect_of_emotions}, the physiological changes in the cardiac system and \glsdesc{eda} are key indicators for affective states.
Therefore, most studies utilise these modalities.
Nevertheless, sensor modalities measuring other physiological parameters (e.g. respiration or muscle activity) might contain valuable information related to an individual's affective state as well.
\autoref{tab:sens_place} lists all relevant sensors and derived indicators, grouped according to their placement on the human body.
This table also provides a minimal recommended sampling rate for each modality in order to being able to extract the respective indicators.
Each of the listed modalities is discussed below, including advantages and limitations with respect to wearable affect recognition.

\begin{table}[t!]
	\centering
	\footnotesize
	\caption{Sensor modalities and derived indicators used in the wearable affect recognition literature.\newline
		\glsreset{hr} \glsreset{hrv} \glsreset{scl} \glsreset{scr}
		\footnotesize{Abbreviations: Minimal required sampling rate in Hz (Min Samp), \gls{hr}, \gls{hrv}}} 
	\label{tab:sens_place}
	\vspace{-1em}
	\begin{tabular}{|l|l|l|l|}
		\hline
		Body Location & Sensor Modality & Min Samp & Derived Indicators \\ \hline
		& \multirow{2}{*}{\Glsdesc{eeg}}
		& \multirow{2}{*}{128}
		& Electric potential changes\\
		&&& of brain neurons\\
		Head/Face	& \Glsdesc{emg}
		& 1000
		& Facial muscle activity\\
		&&&(e.g. zygomaticus major)\\
		& \Glsdesc{eog}
		& 128
		& Eye movements\\ 
		& \Glsdesc{ppg} (ear)
		& 50
		& \gls{hr} and \gls{hrv}\\
		\hline		
		
		& \Glsdesc{ecg}
		& 50
		& \gls{hr} and \gls{hrv}\\  
		& \Glsdesc{eda}
		& 31
		& Tonic and phasic component\\
		Torso			& \Glsdesc{emg}
		& 1000
		& Muscle activity\\ 
		&&&(e.g. upper trapezius muscle)\\
		&Inertial sensor
		& 32
		& Physical activity and body pose\\ 
		& \multirow{2}{*}{\Glsdesc{rip}}
		& \multirow{2}{*}{31}
		& Respiration rate and volume \\
		&			
		&&(thoracic or abdominal)\\
		& Body thermometer
		&1
		& Temperature\\ 
		\hline 
		
		& \Glsdesc{eda} meter
		& 31
		& Tonic and phasic component\\
		& Blood Oxymeter
		&1
		& Blood oxygen saturation\\ 
		Hand/Wrist	& Blood pressure
		&
		& Sphygmomanometer\\	
		& Inertial sensor
		& 32
		& Physical activity\\ 
		& \Glsdesc{ppg}
		& 50
		& \gls{hr} and \gls{hrv}\\ 
		& Thermometer
		& 1
		& Temperature \\	
		\hline 
		
		\multirow{2}{*}{Feet/Ankle}
		& \Glsdesc{eda}
		& 30
		& Tonic and phasic component\\	
		& Inertial sensor
		& 32
		& Physical activity\\ 
		\hline \hline
		
		\multirow{2}{*}{Context}& Sensors of a mobile phone & & Geographic location, Ambient sound, \\
		& (GPS, microphone, etc.)		&		& Physical activity, Social interaction \\
		\hline
	\end{tabular}
	\label{tab:sens}
	\vspace{-1em}
\end{table}

\glsreset{hr} \glsreset{hrv}
In order to assess the \gls{hr}, \gls{hrv} and other parameters related to the cardiac cycle, the \textbf{\gls{ecg}} serves as gold standard.
For a standard three-point \gls{ecg}, three electrodes are placed on the subject's torso, measuring the depolarisation and repolarisation of the heart tissue during each heartbeat.
Sampling rates of \gls{ecg} devices range up to 1024 Hz.
However, when acquired with such high frequency the signal can be downsampled to 256 Hz without loss of information\cite{2012b_Soleymani}.
Furthermore, experiments of \citet{2015_Mahdiani} indicate that a 50 Hz  \gls{ecg} sampling rate is sufficient to obtain \gls{hrv}-related parameters with a reasonable error.
Using \textbf{\gls{ppg}} also provides information about the cardiac cycles.
The \gls{ppg} modality utilises an optical method: The skin voxel, beneath the sensor, is illuminated by a LED of certain wavelength (e.g. green).
A  photodiode measures the amount of backscattered light, and each cardiac cycle appears as a peak in the \gls{ppg} signal.
Data obtained from a \gls{ppg} sensor tends to be much noisier than \gls{ecg} data.
This is due to artefacts caused by e.g. motion, light from external sources, or different tans, which influence the reflection/absorption properties of the skin.
\Gls{ppg} sensors can be attached to the ear, wrist\cite{2017a_Gjoreski} or the finger tip\cite{2014_Lin} of subjects.
The \gls{ppg} modality finds broad application in fitness trackers and smart watches, which can be attributed to the small form factor of the sensory setup. 
Typical sampling rates of \gls{ppg} devices are below 100 Hz.

\glsreset{eda}
The \textbf{\gls{eda}} is commonly measured at locations with a high density of sweat glands, e.g. palm/finger\cite{2012_Choi} or feet\cite{2005_Healey}.
Alternative locations to measure an \gls{eda} signal are the wrist\cite{2017a_Gjoreski} or the torso of a subject\cite{2018_Schmidt}.
In order to asses \gls{eda}, the resistance between two electrodes is measured.
From a technical point of view, the \gls{eda} can be measured using either a constant-voltage or a constant-current system.
Due to practical reasons, the constant voltage-systems find broader application\cite{2000_Dawson}.
\citet{2005_Healey} performed startle measurements on an \gls{eda} signal acquired at 31 Hz. 
This indicates that the minimal sampling rate to decompose the \gls{eda} signal into skin conductance level and skin conductance response contributions is around 30 Hz.
\Gls{eda} electrodes can easily be integrated into wearable systems (e.g. smart watches), allowing minimally intrusive measurement.
However, the \gls{eda} signal is affected by external parameters such as the physical activity of the user, as well as humidity or temperature.

\glsreset{rip}
Although respiration can be assessed indirectly from measuring the blood oxygen level, a direct measurement contains more information about the actual respiration pattern.
Commonly a chest belt (\textbf{\gls{rip}}\cite{2011_Plarre}), which is either worn thoracically or abdominally, is utilised to measure the respiration pattern directly.
During a respiration cycle (inhalation and exhalation), the thorax expands and constricts.
Hence, the chest belt experiences a sinusoidal stretching and destretching process, from which different physiological parameters like respiration rate and volume can be derived.
\citet{2005_Healey} sampled their respiration sensor at 31 Hz, which can be seen as lower boundary of the sampling rate.

\glsreset{emg}
\glsreset{ar}
Muscle activity is measured using surface \textbf{\gls{emg}}, which detects electrical potential.
For this purpose, a pair (or array) of electrodes is attached to the skin above the muscle under consideration.
The electrical potential is generated when the muscle cells are activated, and the surface electrodes are used to recorded this potential difference.
The frequency range of the muscle activity ranges from 15 to 500 Hz\cite{2001_Boxtel}.
Hence, in accordance with the Nyquist theorem, the minimal sampling rate of the \gls{emg} modality should be around 1000 Hz.
One source of noise in surface \gls{emg} are potential changes in adjacent muscles and heart rate activities.
Depending on the measurement position, the QRS complex of the heart can cause artefacts which require additional postprocessing beyond normal filtering.
Considering related work in the affect recognition literature, the \gls{emg} electrodes are commonly placed in the face (e.g. on the zygomaticus major\cite{2012_Koelstra}) or on the shoulder (e.g. on the upper trapezius muscle\cite{2008_AKim, 2010_Wijsman, 2012_Koelstra}).

\glsreset{skt}
Changes in the \textbf{\gls{skt}} can be an indicator for the 'fight or flight' response.
During this response the blood flow to the extremities is restricted, in favour of an increased blood supply of the vital organs.
As a result, the temperature of the extremities decreases.
The temperature can be measured using an infrared thermopile or a temperature-dependent resistor.
As changes of the body temperature are low-frequent, a sampling rate of 1 Hz is sufficient.

\glsreset{eeg} \glsreset{eog} \glsreset{pd}
The physiological modalities detailed above are only minimally intrusive.
Hence, they are frequently employed in affective computing lab and even field studies\cite{2004_Lisetti, 2012_Choi, 2005_Healey,2004_Kim}.
In contrast, \textbf{\gls{eeg}} (which measures the ionic current of brain neurons using electrodes placed on the scalp\cite{2012_Soleymani}), and \textbf{\gls{eog}} (which measures the horizontal and vertical eye movements by placing electrodes above/below and left/right of the eye\cite{2012_Koelstra}) are quite intrusive in our opinion.
Moreover, these modalities are prone to noise generated by muscle activity, posing strong limitations to applicability in real-life scenarios.
Therefore, the modalities EEG and EOG will be given less attention here and in the remainder of this paper.

\glsreset{acc}
\textbf{Inertial sensors}, incorporating a \gls{acc}, gyroscope, and magnetometer, are commonly used in \glsdesc{har}.	
In \glsdesc{ar} field studies the \gls{acc} signal is used to provide context information about the physical activity of the user.
\citet{2017a_Gjoreski}, for instance, used the \gls{acc} signal of a wrist-worn device to classify six different activity types (\textit{lying, sitting, standing, walking, running}, and \textit{cycling}).
The activity, was then used as a feature for their stress classifying system.
However, the results of \citet{2014_Ramos} indicate that in order to detect stress it is sufficient to estimate the intensity level of an activity instead of performing an exact activity classification.

\begin{table}[H]
	\centering
	\footnotesize
	\caption{Overview of the analysed studies from the wearable affect and stress recognition literature.\newline
		\footnotesize{Abbreviations: \glsdesc{acc} (\gls{acc}), \glsdesc{bp} (\gls{bp}), \glsdesc{ecg} (\gls{ecg}), \glsdesc{eda} (\gls{eda}), \glsdesc{eeg} (\gls{eeg}), \glsdesc{emg} (\gls{emg}), \glsdesc{eog} (\gls{eog}), \glsdesc{hr} (\gls{hr}), magnetoencephalogram (MEG), \glsdesc{pd} (\gls{pd}), \glsdesc{ppg} (\gls{ppg}), \glsdesc{resp} (\gls{resp}), \glsdesc{skt} (\gls{skt}), arterial oxygen level (SpO2)}}
	\vspace{-1em}
	\begin{tabular}{
			|>{\raggedright}p{2cm}
			|>{\raggedright}p{7.5cm}
			|>{\raggedright}p{3.5cm}|
		}
		\hline
		Author & Affective States & Included Sensor Modalities\tabularnewline \hline
		\citeauthor{2001_Picard}
		&Neutral, anger, hate, grief, joy, platonic/romantic love, reverence&
		\gls{eda}, \gls{emg}, \gls{ppg}, \gls{resp}
		\tabularnewline \hline
		
		\citeauthor{2004_Haag}
		&Low/medium/high arousal and positive/negative valence&
		\gls{ecg}, \gls{eda}, \gls{emg}, \gls{skt}, \gls{ppg}, \gls{resp}
		\tabularnewline \hline 
		
		\citeauthor{2004_Lisetti}
		&Sadness, anger, fear, surprise, frustration, amusement&
		\gls{ecg}, \gls{eda}, \gls{skt}
		\tabularnewline \hline
		
		\citeauthor{2005_Liu}
		&Anxiety, boredom, engagement, frustration, anger&
		\gls{ecg}, \gls{eda}, \gls{emg}
		\tabularnewline \hline

		\citeauthor{2005_Wagner}
		&Joy, anger, pleasure, sadness & 
		\gls{ecg}, \gls{eda}, \gls{emg}, \gls{resp}
		\tabularnewline \hline
		
		\citeauthor{2005_Healey}
		&Three stress levels&
		\gls{ecg}, \gls{eda}, \gls{emg}, \gls{resp}
		\tabularnewline \hline
		
		\citeauthor{2007_Leon}
		&Neutral/positive/negative valence&
		\gls{eda}, \gls{hr}, BP
		\tabularnewline\hline
		
		\citeauthor{2008_Zhai}
		&Relaxed and stressed&
		\gls{eda}, \gls{pd}, \gls{ppg}, \gls{skt}
		\tabularnewline \hline
		
		\citeauthor{2008_Kim}
		&Distinguish high/low stress group of individuals&
		\gls{ppg}
		\tabularnewline\hline
		
		\citeauthor{2008_AKim}&
		Four quadrants in valence-arousal space&\gls{ecg}, \gls{eda}, \gls{emg}, \gls{resp}
		\tabularnewline\hline
		
		\citeauthor{2008_Katsis}
		&High stress, low stress, disappointment, euphoria&
		\gls{ecg}, \gls{eda}, \gls{emg}, \gls{resp}
		\tabularnewline\hline
		
		\citeauthor{2009_Calvo}
		&Neutral, anger, hate, grief, joy, platonic/romantic love, reverence& \gls{ecg}, \gls{emg}
		\tabularnewline\hline
		
		\multirow{ 2}{*}{\citeauthor{2009_Channel}}	
		&Positively/negatively excited, calm-neutral (in valence-arousal space)&
		\gls{bp}, \gls{eeg}, \gls{eda}, \gls{ppg}, \gls{resp}
		\tabularnewline\hline

		\citeauthor{2009_Khalili}
		&Positively/negatively excited, calm (in valence-arousal space)&
		\gls{bp}, \gls{eeg}, \gls{eda}, \gls{resp}, \gls{skt}
		\tabularnewline\hline
		
		\citeauthor{2010_Healey}
		&Points in valence arousal space. moods& \gls{acc}, \gls{eda}, \gls{hr}, audio
		\tabularnewline\hline
		
		\multirow{ 2}{*}{\citeauthor{2011_Plarre}}
		&Baseline, different types of stress (social, cognitive, and physical), perceived stress&
		\gls{acc}, \gls{ecg}, \gls{eda}, \gls{resp}, \gls{skt}, ambient temperature
		\tabularnewline\hline
		
		\citeauthor{2011_Hernandez}
		&Detect stressful calls&
		\gls{eda}
		\tabularnewline\hline
		
		\citeauthor{2012_Valenza}
		&Five classes of arousal and five valence levels&\gls{ecg}, \gls{eda}, \gls{resp}
		\tabularnewline \hline
		
		\citeauthor{2012_Hamdi}
		&Joy, sadness, disgust, anger, fear, surprise& \gls{ecg}, \gls{eeg}, \gls{emg}
		\tabularnewline\hline				
		
		\multirow{ 2}{*}{\citeauthor{2012_Agrafioti}}
		&Neutral, gore, fear, disgust, excitement, erotica, game elicited mental arousal
		&\multirow{ 2}{*}{\gls{ecg}}\tabularnewline\hline
		
		\multirow{ 2}{*}{\citeauthor{2012_Koelstra}}
		& \multirow{ 2}{*}{Four quadrants in valence-arousal space}&\gls{ecg}, \gls{eda}, \gls{eeg}, \gls{emg}, \gls{eog}, \gls{resp}, \gls{skt}, facial video \tabularnewline \hline
		
		\multirow{ 2}{*}{\citeauthor{2012b_Soleymani}}
		&Neutral, anxiety, amusement, sadness, joy, disgust, anger, surprise, fear&
		\multirow{ 2}{*}{\gls{ecg}, \gls{eda}, \gls{eeg}, \gls{resp}, \gls{skt}}
		\tabularnewline\hline
		
		\citeauthor{2013_Sano}
		&Stress vs. neutral &\gls{acc}, \gls{eda}, phone usage
		\tabularnewline\hline
		
		\citeauthor{2013_Martinez}
		&Relaxation, anxiety, excitement, fun&
		\gls{eda}, \gls{ppg}\tabularnewline \hline
		
		\citeauthor{2014_Valenza}
		&Four quadrants in valence-arousal space &\gls{ecg}
		\tabularnewline\hline
		
		\citeauthor{2014_Adams}
		&Stress vs. neutral (aroused vs. non-aroused)&\gls{eda}, audio
		\tabularnewline\hline
		
		\citeauthor{2015_Hovsepian}					
		&Stress vs. neutral & \gls{ecg}, \gls{resp}
		\tabularnewline\hline
		
		\multirow{ 2}{*}{\citeauthor{2015_Abadi}}
		&\multirow{ 2}{*}{High/Low valence, arousal, and dominance} &\gls{ecg}, \gls{eog}, \gls{emg}, near-infrared face video, MEG
		\tabularnewline \hline
		
		\citeauthor{2016_Rubin}
		& Panic attack &\gls{acc}, \gls{ecg}, \gls{resp}
		\tabularnewline \hline
		
		\citeauthor{2016_Jaques}
		&Stress, happiness, health values&\gls{eda},\gls{skt}, \gls{acc}, phone usage
		\tabularnewline\hline
		
		\citeauthor{2016_Rathod}
		&Normal, happy, sad, fear, anger&\gls{eda}, \gls{ppg}
		\tabularnewline \hline
		
		\citeauthor{2016_Zenonos}
		&Excited, happy, calm, tired, bored, sad, stressed, angry&  \gls{acc}, \gls{ecg}, \gls{ppg}, \gls{skt}
		\tabularnewline\hline
		
		\citeauthor{2016_Zhu}
		&Angle in valence arousal space&\gls{acc}, phone context
		\tabularnewline \hline
		
		\citeauthor{2016_Birjandtalab}
		&Relaxation, different types of stress (physical, emotional, cognitive)&\gls{acc}, \gls{eda}, \gls{skt}, \gls{hr}, SpO2
		\tabularnewline\hline
		
		\citeauthor{2017a_Gjoreski}	& Lab: no/low/high stress; Field: stress vs. neutral&\gls{acc}, \gls{eda}, \gls{ppg}, \gls{skt}
		\tabularnewline \hline
		
		\citeauthor{2017_Mozos} &Stress vs. neutral& \gls{acc}, \gls{eda}, \gls{ppg}, audio
		\tabularnewline \hline
		
		\multirow{ 3}{*}{\citeauthor{2018_Schmidt}}
		& 	
		\multirow{ 3}{*}{Neutral, fun, stress} & Torso: \gls{acc}, \gls{ecg}, \gls{eda}, \gls{emg}, \gls{resp}, \gls{skt}; \\ Wrist: \gls{acc}, \gls{eda}, \gls{ppg}, \gls{skt},
		\tabularnewline \hline
		
	\end{tabular}
	\label{tab:aim}
\end{table}

Finally, following \citet{2013_Muaremi}, (smart) phones offer an ideal platform to generate additional \textbf{contextual data}.
This contextual data is aggregated by utilising position (GPS), sound snippets, calender events, ambient light, and user interaction with the phone\cite{2013_Muaremi,2017_Mozos}.

\subsection{Affect and stress recognition studies based on wearable sensor data}
\label{sub:freq_em_sensors}

\glsreset{ar}
This section summarises studies aiming to detect different affective states, based on wearable sensor data.
In \autoref{tab:aim}, the target affective states and the employed sensor modalities of the studies are detailed. 
Later, in \Cref{sub:class}, a comprehensive comparison of these studies will be presented, focusing on the employed classification algorithms and achieved performance (see \autoref{tab:lit_meth}).

The affective states to be recognised are rather diverse in the studies presented in \autoref{tab:aim}.
However, almost 38\% of the studies aimed to detect stress.
For this purpose, different types of stressors (e.g. mental, physical or social\cite{2011_Plarre, 2016_Birjandtalab}) or different stress levels\cite{2017a_Gjoreski} are distinguished. 
The popularity of these automated stress detection systems clearly stem from the fact that stress recognition is highly relevant from a healthcare point of view (see \Cref{sec:intro}).
According to \autoref{tab:aim}, various studies aim to recognise different emotional categories and distinguish between up to eight different affective states.
Finally, dimensional representations of emotions (mainly relying on the valence-arousal space) were used in 27\% of the analysed studies.

Concluding from \autoref{tab:aim}, sensor modalities monitoring the cardiac cycle (e.g. \gls{ecg} or \gls{ppg}) are employed in 78\% of the studies. 
The \glsdesc{eda} was recorded in 76\% of the studies and is the second most frequently used modality.
Arousal-related changes in the affective state are known to have an impact on the cardiac cycle and the sweat production (see \Cref{sub:af_foot}). 
Hence, the popularity of these modalities is easily explained.
In 40\% of the studies presented in \autoref{tab:aim}, respiration data was acquired.
\citet{2008_AKim} pointed out that increased arousal can lead to irregular respiration patter.
Finally, skin temperature, 3-axis acceleration, and electromyogram data were recorded in 32\% of the studies.

\section{Data generation protocols}
\label{sec:data_gt}

\citet{2001_Picard} pointed out that, in order to generate high quality physiological data for affect detection, carefully designed study protocols are required. 
The arguably most important decision is whether the experiment is to be conducted in a laboratory setting or in the natural environment of the subjects.
A key issue when designing a field study is the accurate label generation, concerning the subjects' affective states.
In contrast, during a lab study, obtaining high quality labels is a minor issue as either the study protocol can be used or dedicated time slots for questionnaires can be reserved.
However, considering lab user studies, the desired affective states have to be elicited by carefully choosing a set of stimuli.
On the other hand, during field studies, affective stimuli do not have to be designed, different affective states occur naturally.
\Cref{sub:lab} provides an overview on data generation protocols for user studies in the lab. \Cref{sub:field} summarises related work on how to plan and conduct affect-related field studies.
Finally, based on the overview given on lab and field studies, we provide practical guidelines for designing and applying questionnaires in \Cref{sub:guide}.

\subsection{Affect-related user studies in laboratory settings}
\label{sub:lab}

Obtaining ground truth in lab studies can be often solved by relying on the study protocol. In addition, questionnaires can be scheduled and integrated into the study protocol to verify that the desired affective states are successfully evoked by obtaining subjective assessment from the study participant. Typically, these questionnaires are used directly after each affective stimulus or condition. For example, \citet{2014_Ramos} collected subjective stress levels after each stressor in their lab study. However, since obtaining accurate labels in field studies is more challenging, a thorough overview on this topic will be given in \Cref{sub:field}. Therefore, the remaining of this section will focus on how to induce different affective states.

Humans differ in their personality and hence generating data that corresponds to a particular emotional state is a challenging task\cite{2012_Hamdi}.
However, due to the controlled environment of a lab study, researchers can conduct studies following well-designed protocols.
These protocols can be tailored to elicit different emotional states in the study participants. 
Moreover, due to the controlled environment, lab studies can be applied in the same way to multiple subjects and can be replicated by other researchers.
Lab study protocols in related work use well-studied and validated stimuli which reliably induce affective states.
In the following we give an overview of these stimuli.

\textbf{Images:} The \gls{iaps}\cite{1999_Lang} is a data set comprised of colour photographs.
The images in IAPS were chosen to elicit emotional reactions.
Each image was rated multiple times by study participants, providing labels in the valence and arousal space. 
\citet{2005_Mikels} created a subset of the IAPS from images identified as eliciting certain discrete emotions.
Hence, depending on the desired emotion, one can choose particularly strong images from this subset.
In the domain of affective computing, the IAPS has for instance been used by \citet{2007_Leon} and by \citet{2012_Hamdi}.
In the experiments presented by \citet{2007_Leon}, 21 images from the IAPS were used to elicit three different affective states (neutral, positive, negative).
\citet{2012_Hamdi} exposed their study participants to ten images from the IAPS and aimed at recognising six basic emotions (disgust, joy, surprise, sadness, fear, anger) based on physiological data. 

\textbf{Videos:} According to \citet{1995_Gross}, short audiovisual clips are very suitable to elicit discrete emotions.
Hence, video clips are frequently employed as stimuli\cite{2012b_Soleymani, 2015_Abadi, 2012_Koelstra}. 
A common procedure to select a set of videos evoking certain target emotions is to chose them from a large pool of videos. 
The process of identifying the 'right' videos is often done in two steps. 
First, the clips are watched and rated by a large number of individuals. 
Second, the clips which elicit a certain emotion\cite{2012_Soleymani, 2012_Koelstra} most reliably are chosen as stimuli in the study.
Recently, \citet{2016_Samson} published a study on 199 amateur clips (each $20$ to $33$ seconds long), which were rated by 411 subjects with respect to three affective categories (neutral, positive, and negative).
In the affect recognition literature, there exist many examples of relying on audiovisual clips to elicit the desired affective states.
\citet{2012_Koelstra} chose in their experiments music clips with a length of 60 seconds.
After each stimulus, the current trial number (2 seconds) and a 5 second baseline were recorded.
\citet{2012_Soleymani} showed their study participants 60 to 120 seconds long excerpts from movies and after each clip a short neutral clip (~15 seconds) was displayed.

\textbf{Acted Emotions:} In the above detailed protocols, emotions are event-elicited.
Another way of generating affective states is to ask the subjects to purposefully elicit emotions, e.g. act an emotion.
For instance, \citet{2017_Hanai} asked the study participants to tell at least one happy and one sad story.
Other researchers\cite{2008_Castellano, 2013_Dobrisek} asked trained actors to perform certain emotions.
These types of approaches are frequently employed in sentiment analysis and emotion recognition from audio/video data.
A popular and freely available data set of acted emotions is IEMOCAP\cite{2008_Busso}.

\textbf{Game elicited emotions}: Another way to elicit a certain affective state is to ask the subject to perform a certain task.
\citet{2015_Taylor} were able to elicit frustration in their study participants by implementing latency between the user's touch and the reaction of the Breakout engine.
\citet{2013_Martinez} used four different versions of a Maze-Ball game to generate pairwise preference scores. The scores were generated by asking the subjects which of two games felt more \textit{anxious, exciting, frustrating, fun} and \textit{relaxing}.

\textbf{Stress inducing study protocols:}
Beside of eliciting certain emotions, there are numerous protocols to generate some sort of stress, especially in laboratory settings.
\citet{1968_Mason} showed that in order to trigger a stress response, the situation has to be either novel, and/or unpredictable, and/or the subject has to have the feeling that he/she is not in control of the situation\cite{2007_Lupien}.
Stressors employed in related work can typically be categorised as:
\begin{enumerate}
	\item Social-evaluative: a task creating a socially relevant situation for the subject. For example, performing a task in front of a panel which evaluates the subject.
	\item Cognitive: a task demanding significant mental engagement and attention. For example, performing an arithmetic task.
	\item Physical: a task creating a physically uncomfortable situation. For example, being exposed to extreme hot or cold.
\end{enumerate}

A well-studied and frequently employed stress elicitation protocol is the \gls{tsst}\cite{1993_Kirschbaum}.
It consists of a public speaking/job interview type of situation (represents the social-evaluative category), and a mental arithmetic task, inflicting high cognitive load.
Due to its reliability and easy set-up, the \gls{tsst} was administered in numerous studies, e.g. \citet{2017_Mozos, 2011_Plarre, 2015_Hovsepian}, and \citet{2016_Gjoreski}.
\citet{2005_Dedovic} modified the \gls{tsst}-protocol so that it can be applied in fMRI studies, giving rise to the \gls{mist}.
Another task inflicting cognitive load is the \textit{Stroop color test}\cite{1935_Stroop}.
Following this protocol, the subjects are presented with the word of a colour displayed in a different colour than its meaning (e.g. \textcolor{blue}{green}), and the task is to read the word out loud.
\citet{2012_Choi}, for instance, employed the Stroop color test in their experiments to develop a wearable stress monitoring system.

Stress can be also elicited using computer tasks.
\citet{2013_Wijsman}, for instance, asked the subjects to perform a calculation task, to solve a logical puzzle, and to do memorisation task.
These tasks had to be completed under time pressure, while exposing the subjects to distracting sounds.
Furthermore, the subjects were filmed during the memorisation task.
In addition, this task had a social-evaluative component as the participants were told that their scores would be available for their colleagues afterwards.
The cold pressor test, applied by \citet{2011_Plarre}, can be used to evoke physical stress. 
Following this test, the subjects are asked to place their (dominant) hand into a bucket of ice cold water and leave it there for a predefined time (e.g. $60$ seconds).

Considering a laboratory setting, another important aspect is to decide to which degree the subjects are to be informed about the protocol and aim of the study.
In order to reduce participants' bias it can be sometimes necessary to disguise the true purpose of the study.
If the study protocol requires to deceive the participants, it is important to get the approval of an ethics committee and to uncover the true aim of the study as soon as the experiment is over.\\


\vspace{-0.5em}
\subsection{Affect-related user studies in the field}
\label{sub:field}

\begin{table}[]
	\centering
	\footnotesize
	\caption{Methods used to generate ground truth in field studies.\newline
		\footnotesize{Abbreviation: \glsdesc{ema} (\gls{ema}), \glsdesc{pss} (\gls{pss}), \glsdesc{stai} (\gls{stai}), \glsdesc{psqi} (\gls{psqi}), \glsdesc{bfi} (\gls{bfi}), \glsdesc{panas} (\gls{panas}), \glsdesc{sri} (\gls{sri})}}
	\vspace{-1em}
	\begin{tabular}{|>{\raggedright}p{0.2cm}
			|>{\raggedright}p{2cm}
			|>{\raggedright}p{10.7cm}|}
		\hline
		& Author & Ground truth generation approach \tabularnewline\hline
		\parbox[t]{2mm}{\multirow{6}{*}{\rotatebox[origin=c]{90}{Emotion}}}
		&\citeauthor{2010_Healey}, \citeyear{2010_Healey}
		&Participants filled in an \gls{ema} whenever they felt a change in their affective/ physiological state. \glspl{ema} included a form of the circumplex model and a field for free text. Interviews at the end of the workday to generate additional labels and revision. \tabularnewline \cline{2-3}
		&\citeauthor{2015_Rubin},\\ \citeyear{2015_Rubin}
		&Start/stop time and rating of 15 panic attack symptoms according to their severity were reported by the subject using a mobile app. \tabularnewline\cline{2-3}
		&\citeauthor{2016_Jaques}, \citeyear{2016_Jaques}
		& Students reported health, stress and happiness twice a day (morning and evening).\tabularnewline \hline \hline
		
		\parbox[t]{2mm}{\multirow{17}{*}{\rotatebox[origin=c]{90}{Stress}}}
		&\citeauthor{2011_Hernandez}, \citeyear{2011_Hernandez}
		&Seven employees of a call center rated all their incoming calls from on a 7 point likert scale.\tabularnewline \cline{2-3}
		&\citeauthor{2013_Muaremi}, \citeyear{2013_Muaremi}
		&Participants were asked to fill in a shortened \gls{panas} four times between 8\,a.m and 8\,p.m. Before going to sleep they reported how stressful they felt during the day. \tabularnewline\cline{2-3}
		&\citeauthor{2008_Kim}, \citeyear{2008_Kim}
		&Pre-study: in order to divide the subjects into two groups they filled out a simplified \gls{sri}.\tabularnewline\cline{2-3}
		&\citeauthor{2013_Sano}, \citeyear{2013_Sano}
		&Pre-study: Participants filled in a \gls{pss}, \gls{psqi}, and \gls{bfi}.\\ 
		During study: morning and evening \glspl{ema} on sleep, mood, stress level, health and other.\\
		Post-study: Participants filled in questionnaires on health, mood, and stress.\tabularnewline \cline{2-3}
		&\citeauthor{2014_Adams},  \citeyear{2014_Adams}
		&Pre-study: Participants filled in a \gls{panas}, \gls{pss} and a measure of mindfulness.\\
		During study: Self-reports approximately every 30 min. (with small random variations). Participants reported on momentary stress and affect. Additional reports and a small free text field were available too.\\
		Post-study: semi-structured interview at the end of the end data collection.\tabularnewline \cline{2-3}
		&\citeauthor{2015_Hovsepian}, \citeyear{2015_Hovsepian}
		&\Glspl{ema} randomly scheduled approximately 15 times. During each \gls{ema} subjects filled in a shortened version of the \gls{pss} containing 6 items.\tabularnewline\cline{2-3}
		&\citeauthor{2016_Gjoreski}, \citeyear{2016_Gjoreski}
		&Subjects replied to 4 to 6 randomly scheduled \glspl{ema}. During each \gls{ema} subjects reported on their current stress level.\tabularnewline\hline\hline
		
		\parbox[t]{2mm}{\multirow{13}{*}{\rotatebox[origin=c]{90}{Mood}}}
		
		&\citeauthor{2013_LiKamWa}, \citeyear{2013_LiKamWa}&Participants were asked to report their mood four times a day. \glspl{ema} consist of two sliders representing the pleasure and activeness dimension of the circumplex model.\tabularnewline \cline{2-3}
		&\citeauthor{2014_Wang},\\ \citeyear{2014_Wang}
		& Pre-study:  Subject filled in a number of behavioural and health surveys.\\
		During study: Every participant fills in 8 \glspl{ema} every day. The \glspl{ema}  include measures on mood, health, stress and other affective states.\\
		Post-study: Interviews and the same set of behavioural and health surveys were administered.\tabularnewline \cline{2-3}
		&\citeauthor{2015_Sano},\\ \citeyear{2015_Sano}&
		Pre-study: subjects filled in \gls{bfi}, \gls{psqi} and the Morningness-Eveningness\cite{1976_Horne} questionnaire.\\
		During study: similar to \citet{2013_Sano} subject filled \glspl{ema} morning and evening on: activities, sleep, social interaction, health,mood, stress level and tiredness.\\
		Post-study: Subjects filed in a \gls{pss}, \gls{stai}, and other questionnaires related to physical and mental health.\tabularnewline \cline{2-3}
		&\citeauthor{2016_Zenonos}, \citeyear{2016_Zenonos}& \glspl{ema} were scheduled every two hours. For the \glspl{ema} a app was used, containing sliders from 0-100 for 8 moods. Additionally, a free text field was provided.\tabularnewline \hline
	\end{tabular}
	\label{tab:motiv}
\end{table}

Data collection in the wild is essential to develop systems designed for everyday use. Different affective states do not have to be induced as they occur naturally. However, since no data collection protocol is available in these scenarios, the availability of ground truth information has to be ensured differently. Thereby, an important trade-off has to be considered: Get informed about the subject's affective state as frequently as possible while not overloading the subject with questionnaires and forcing to interrupt daily routine too often.
\autoref{tab:motiv} gives an overview on field studies carried out within the scope of wearable affect recognition.
The focus of \autoref{tab:motiv} is on the methods used to obtain ground truth, summarising the affective states of interest and the type and scheduling of the applied questionnaires.
Further details to each of the used questionnaires will be given in \autoref{tab:ema_list}.

Most commonly, ground truth collection in the wild relies on the so-called \gls{ema}, also referred to as the experience sampling method. The idea of this method is that subjects record their emotions or mood in the moment, typically via self-reports. Participants of field studies are either prompted to complete such self-reports at certain times over a day or prompts are event-triggered. For example, \citet{2016_Zenonos} conducted a study on employees during their work hours with self-reports every two hours. The EMAs in this study refer to eight different moods, asking for each the question \textit{How have you been feeling for the last two hours?} Another approach is to distribute a defined number of EMAs randomly over the day. This way, subjects are not prepared to the self-report requests.
For example, \citet{2013_Muaremi} divided the day into four sections, and asked subjects randomly within each section to fill in a self-report.
In case a user study focuses on specific affective states, an event-triggered self-reporting approach can be applied. For example, in the study of \citet{2011_Hernandez} call centre employees rated personal stress level after each call, or in the study of \citet{2015_Rubin} self-reports were filled out when subjects became aware of the symptoms of a panic attack. In addition to EMAs (either scheduled or event-triggered), a few studies also performed directed interviews. For example, \citet{2010_Healey} conducted daily interviews at the end of each work day, with the goal to understand and correct/extend participants' annotations.
\begin{table}[t!]
	\footnotesize
	\centering
	\caption{Standard and self-defined EMAs applied in user studies of wearable affect recognition.}
	\vspace{-1em}
	\begin{tabular}
		{|>{\raggedright}p{3cm}
			|>{\raggedright}p{4cm}
			|>{\centering}p{1cm}
			|>{\centering}p{2cm}
			|>{\centering}p{2.0cm}|}
		\hline
		Measurement goal & EMA-tool and description & Items & Source & Example use \tabularnewline
		\hline
		Affective state as a point in valence-arousal space & Mood Map: a translation of the circumplex model of emotion & 
		2& \citet{2009_Morris} & \citet{2010_Healey} \tabularnewline 
		\cline{2-5}
		& Mood Journal: 5-point scale for pleasure and activeness dimensions &
		2 & \citet{2013_LiKamWa}  & \citet{2013_LiKamWa} \tabularnewline
		\hline
		Measure positive and negative affect & \glsdesc{panas} (\gls{panas})questionnaire& 
		20 & \citet{1988_Watson} & \citet{2014_Adams} \tabularnewline
		\cline{2-5}
		& Shortened \gls{panas}& 
		10& \citet{2013_Muaremi} &\citet{2013_Muaremi} \tabularnewline
		\cline{2-5}
		Estimate PA part of the \gls{panas} & PAM (Photo Affect Meter): choose one of 16 images, mapped to the valence-arousal space &
		1 & \citet{2011_Pollak} & \citet{2014_Wang}\tabularnewline
		\hline
		Measure success areas like relationships, self-esteem, purpose, and optimism & Flourishing scale & 
		8&\citet{2010_Diener} & \citet{2014_Wang} \tabularnewline
		\hline
		Subjective mood indicator & Smartphone app querying user's mood &
		8 & HealthyOffice app\cite{HealthyOff}& \citet{2016_Zenonos} \tabularnewline
		\hline
		Subjective stress level assessment & \glsdesc{pss} (\gls{pss}): subject's perception and awareness of stress &
		14 &\citet{1983_Cohen} & \citet{2013_Sano},\\ \citet{2014_Wang} \tabularnewline
		\cline{2-5}
		& Shortened \gls{pss} used in ambulatory setting
		& 5 & \citet{2015_Hovsepian} & \citet{2015_Hovsepian},\\ \citet{2011_Plarre}\tabularnewline
		\cline{2-5}
		& \gls{sri}: score the severity of stress-related symptoms experienced within a time interval 
		& 22 & \citet{2001_Koh}, \citet{2006_Choi} & \citet{2008_Kim} \tabularnewline
		\hline
		Measure anxiety level& \glsdesc{stai} (\gls{stai}) & 
		20 & \citet{1970_Spielberger} & \citet{2016_Gjoreski} \tabularnewline
		\hline
		Depression level & \glsdesc{phq} (\gls{phq}): score DSM-IV manual &
		9 & \citet{2001_Kroenke}& \citet{2014_Wang} \tabularnewline
		\hline
		Loneliness level & UCLA loneliness scale addressing subjective feelings of loneliness and feelings of social isolation.
		&
		20& \citet{1996_Russel} & \citet{2014_Wang} \tabularnewline
		\hline
		Severity of panic attack symptoms & Symptoms from the DSM-IV and Panic Disorder Severity Scale
		standard instrument&
		15& \citet{1997_Shear}  & \citet{2015_Rubin} \tabularnewline
		\hline
		Sleep behaviour and quality & \glsdesc{psqi} (\gls{psqi}) &
		19&\citet{1989_Buysse} & \citet{2013_Sano} \tabularnewline
		\hline
		Personality, Big Five personality traits & \glsdesc{bfi} (\gls{bfi}) & 44 & \citet{1999_John} & \citet{2013_Sano}, \\\citet{2015_Sano} \tabularnewline
		\cline{2-5}
		& \gls{bfms} & 50 & \citet{2002_Perugini} & \citet{2016_Subramanian} \tabularnewline
		\hline
	\end{tabular}
	\label{tab:ema_list}
\end{table}

Beside the frequency of EMAs over a day, the length and complexity of each single questionnaire are also important factors defining the load on the subjects. In order to avoid overloading study participants, EMAs should be focused on the goal of the study and answering them should be as easy as possible for participants. \citet{2013_Muaremi} for instance report that they received complaints about the difficulty of completing a self-assessment questionnaire which originally consisted of 20 items, and thus reduced the questionnaire to only ten items.
Field studies of wearable affect recognition typically rely on standard questionnaires or their variation (see \autoref{tab:ema_list} for an overview).
An important goal of affect-related studies is to assess a subject's affective state within the valence-arousal space. \citet{2010_Healey} used a tool called Mood Map as EMA, while \citet{2013_LiKamWa} defined a 5-point scale for each dimension as self-report questionnaire. In case only the valence (positive-negative affect) is of interest, the Positive and Negative Affect Schedule (PANAS) questionnaire is a suitable tool. \citet{2014_Adams} used the PANAS questionnaire prior to the study period. \citet{2013_Muaremi} used a shortened version of PANAS during their field study, which consisted of five positive affect items (relaxed, happy, concentrated, interested, and active) and five negative affect items (tired, stressed, sleepy, angry, and depressed). Furthermore, \citet{2014_Wang} applied \gls{pam} as the EMA-tool, in which users select from different images the one which best suits their current affective state. In order to assess the flourishing level of subjects, \citet{2014_Wang} included a flourishing scale EMA in their study. \citet{2016_Zenonos}provide an example for a custom EMA for overall mood assessment: Focusing on eight moods, they asked participants to rate each on a scale from 0-100, generating ground truth information this way.

Focusing on the task of recognising and assessing stress level, several standard psychological questionnaires can be applied for EMAs. The \gls{pss} measures the subject's perception and awareness of stress, consisting of 10 or 14 items. \gls{pss} was used by \citet{2013_Sano} and \citet{2014_Wang} in their respective field studies, while \citet{2015_Hovsepian} used an adapted PSS for ambulatory setting with only five items. The severity of stress-related symptoms can be scored using the \gls{sri}, or a simplified version of it, as shown by \citet{2008_Kim}. In order to measure anxiety, the \gls{stai}  can be used, as shown by \citet{2016_Gjoreski}. EMAs based on standardised questionnaires for depression level (e.g. \gls{phq}) and loneliness level (UCLA loneliness scale) are used by \citet{2014_Wang}. For assessing the severity of panic attacks, \citet{2015_Rubin} created a questionnaire including 15 panic attack symptoms. In case a panic attack occurred, subjects were asked to rate the severity of each of the 15 symptoms, using a severity rating of 1 (none) to 5 (extreme). Gathering information about subjects' sleep quality might be useful in affect-related studies. The \gls{psqi} can serve as a suitable EMA for sleep behaviour and quality assessment, as demonstrated by \citet{2013_Sano}. Finally, subjects' personality traits can have an influence on their affective perception and physiological response\cite{2016_Subramanian}. Therefore, completing one personality-related questionnaire, e.g. at the beginning of a field study, might provide valuable insights into the subjects. For example, \citet{2015_Sano} used the Big Five personality traits as features for predicting subjects' mood.

Historically, personal notebooks or journals were used for EMAs. However, these tools have been predominantly replaced by smartphones, as they offer an ideal platform to facilitate self-reports: Subjects do not need to carry a study-specific device with themselves, EMAs can be automatically scheduled and uploaded (thus, the study supervisor is constantly updated), and contextual information available on the smartphone can be logged together with the generated ground truth information. Moreover, monitoring the participant's motivation in filling out and submitting the self-reports (considering both frequency and completeness) might be important to ensure high-quality labels. In order to increase the participant's motivation, a reward system can be established. For example, participants of the study conducted by \citet{2010_Healey} received a base reward and an incremental reward, depending on the number of annotations made over the day.
Another reward structure was introduced by \citet{2014_Wang}: They offered all subjects a base reward, and the participants providing the most EMA data had the chance to win additional prizes.  

\subsection{Practical guidelines for \glsdesc{ema}}
\label{sub:guide}

Based on the overview given above on lab and field studies, we now provide the reader practical guidelines for designing and applying \glspl{ema}.

\begin{enumerate}
	\item \textbf{Sampling rate}: When defining the number of \glspl{ema} over the observation period, the following trade-off should be considered: Getting informed about the subject's affective state as frequently as possible while not overloading the subject with questionnaires and forcing to interrupt daily routine too often. A good compromise is to schedule an \gls{ema} every two hours\cite{2016_Zenonos} or approximately five times over the day\cite{2016_Gjoreski}.

	\item \textbf{General scheduling}: A good practice is to schedule \glspl{ema} randomly during a field study, in order that subjects are unprepared. In case an approximately evenly distributed scheduling over the observation period is desired, the following approach could be used: Divide the observation period into $N$ sections (where $N$ is the total number of \glspl{ema} over the observation period), and randomly schedule one \gls{ema} within each section. This approach was applied for example by \citet{2013_Muaremi}. Considering user studies in the lab, \glspl{ema} are typically scheduled directly after each affective stimulus or condition.

	\item \textbf{Scheduling by subjects}: As \glspl{ema} are commonly scheduled randomly during field studies, these questionnaires are independent of the study participants' affective states. Therefore, it is good practice to allow subjects to start an \gls{ema} (in addition to the generally scheduled ones) whenever they feel a change in their affective state. For example, \citet{2017a_Gjoreski} enabled their study participants to log stressful events whenever they occurred.
	
	\item \textbf{Number of items}: In order to avoid overloading subjects during a field study, the time required to answer an \gls{ema} should be minimised. Therefore, \glspl{ema} should be focused only on the goal of the study and include as few items to answer as possible. A good compromise is to include at most ten items per scheduled \gls{ema}, as discussed by \citet{2013_Muaremi}. Considering lab studies, the length of an \gls{ema} is usually not a critical issue. Moreover, \glspl{ema} in laboratory settings can be used during the cool-down phase after an affective stimulus, allowing enough time to complete even long \glspl{ema}.
	
	\item \textbf{Situation labels}: It is important to generate labels on the spot and not on hindsight, as a subject's memory can be altered due to memorisation effects (e.g. halo effect: the occurrence of a certain emotion can influence the perception of other affective states experienced during the observation period). Nevertheless, during field studies, it is good practice to review the labels together with the study participant, e.g. on a daily basis.
	
	\item \textbf{Length of labels}: For a (mentally) healthy subject, affective states are expected to be stable on short time scales. However, when labels are generated using \glspl{ema}, the question arises how long these labels are valid. Considering lab studies, labels usually refer to the preceding affective condition. Considering field studies, however, the validity of labels is not as trivial. Depending on the focus of the study, one has to decide on a label length. If the study addresses mood, longer label periods, e.g. 2 hours\cite{2016_Zenonos}, can be taken into account. If the study addresses shorter affective states (emotions or stress), shorter label periods are used. For example, in order to detect and classify stress, \citet{2017a_Gjoreski} considered ten minutes before and after each provided label.

	\item \textbf{Ensure engagement}: Especially considering field studies, it is important to keep subjects motivated to fill out \glspl{ema}. For example, including images in the \gls{ema} (e.g. incorporating \gls{pam} as one item) could make answering questionnaires more interesting. In addition, a common practice is to introduce an incremental reward system in field studies. Overall, keeping the subjects motivated will ensure high-quality labels, regarding both frequency and completeness.
	
\end{enumerate}

\section{Publicly available datasets}
\label{sec:data_set}
Conducting a user study is a time consuming and challenging task, especially if one is interested in real-life data.
Hence, publicly available datasets should be considered.
Furthermore, these datasets can be used for benchmarking algorithms and facilitate a direct comparison of different results.
However, currently only a few publicly available datasets exist which contain solely wearable sensor data of various affective states.
Therefore, we extend the scope of this section to datasets with a broader relevance to wearable affect recognition.
All datasets are included which a) are publicly available, b) include data recorded from study participants being subject to either an emotional stimuli or a stressor, and c) include at least a few sensor modalities which can be (theoretically) worn.
The datasets included in our analysis are listed in \autoref{tab:avail_data}, and are further detailed below.

\textbf{Eight-Emotion} dataset\cite{2001_Picard}: includes data of one (female) study participant who is subject to the same set of stimuli over a time span of 20 days.
The stimuli, a set of personally-significant imagery, were chosen by the subject to elicit the affective states \textit{neutral,
anger, hate, grief, platonic love, romantic love, joy,} and \textit{reverence}. 
The physiological signals (\gls{ecg}, \gls{eda}, \gls{emg}, and \gls{resp}) were sampled at 20 Hz.
Major limitations of this dataset are: a) only one subject is included, and b) aliasing artefacts are likely to occur (especially in the \gls{emg} signal) due to the low sampling rate.

\textbf{DEAP} (Database for Emotion Analysis using Physiological signals)\cite{2012_Koelstra} features physiological data of 32 study participants.
In the study protocol of DEAP, one minute excerpts of music videos were used as stimuli.
These 40 clips were selected from a larger pool according to valence, arousal, and dominance ratings gathered during a pre-study. 
The physiological signals were all sampled with 512 Hz and later downsampled to 256 Hz.
As DEAP includes subjects' ratings of the videos (valence, arousal, dominance, and liking) the dataset can be used to develop models recognising these labels.
However, it is important to note that, due to the sensor setup defined by the DEAP protocol, the study participants were very limited in terms of movement. Therefore, one can expect that models trained on the DEAP dataset will have a limited performance in real-life settings.

\textbf{MAHNOB-HCI}\cite{2012b_Soleymani} includes physiological data from 27 study participants (16 female). Moreover, the dataset includes face and body video from six cameras, and data from an eye gaze tracker and audio.
The  physiological data was sampled at 1024 Hz and later downsampled to 256 Hz.
\citet{2012b_Soleymani} report on two experiments, both included in the MAHNOB-HCI dataset. During the first experiment, study participants watched a set of 20 video clips, each associated with certain emotional keywords (\textit{disgust, amusement, joy, fear, sadness}, and \textit{neutral}).
The goal of the second experiment was implicit tagging: Subjects were exposed to 28 images and 14 videos, and reported on the agreement with the displayed tags.
Considering the topic of affect recognition, data in particular of the first experiment is of interest.

\textbf{StudentLife}\cite{2014_Wang} is a dataset recorded from 48 students (10 female) at the Dartmouth college. 
Each student was monitored over a period of 10 weeks (one academic semester).
As the StudentLife dataset was recorded in the field, no specific stimulus was used to trigger certain affective states.
However, due to the progress of the semester, it is expected that the students were more stressed toward the end of the data collection period (due to the upcoming examination phase of the semester).
StudentLife only contains data recorded from the students' smartphones: data from accelerometer, microphone, light sensor, and GPS/Bluetooth were extracted.
Moreover, various information related to the students' context (e.g. class attendance) and smartphone usage (e.g. conversation frequency and duration) were recorded.
In addition, the dataset includes a large amount of self-reports, addressing physical activity and sleep behaviour, perceived stress, mood, mental well-being, etc.
Due to the popularity of smartphones, the dataset is certainly of interest by facilitating affect and stress recognition purely based on smartphone usage patterns.

\begin{table}[t]
	\centering
	\footnotesize
	\caption{Publicly available datasets relevant for wearable affect and stress recognition.
		\footnotesize{Abbreviations: Number of subjects (Sub), Location (Loc), Lab (L), Field (F), Field with constraint (FC), \glsdesc{acc} (\gls{acc}), arterial oxygen level (SpO2), \glsdesc{ecg} (\gls{ecg}), \glsdesc{eda} (\gls{eda}), \glsdesc{eeg} (\gls{eeg}), \glsdesc{emg} (\gls{emg}), \glsdesc{eog} (\gls{eog}), magnetoencephalogram (MEG), \glsdesc{resp} (\gls{resp}), \glsdesc{skt} (\gls{skt})}}
	\vspace{-1em}
	\begin{tabular}{
			|>{\raggedright}p{.2cm}
			|>{\raggedright}p{1.5cm}
			|>{\raggedright}p{5cm}
			|>{\centering}p{0.5cm}
			|>{\centering}p{0.5cm}
			|>{\raggedright}p{4.3cm}|}\hline
		&Name			& Labels		& Sub.		&Loc.& Included Modalities  
		\tabularnewline \hline
		
		\parbox[t]{2mm}{\multirow{15}{*}{\rotatebox[origin=c]{90}{Emotion (E)}}} 
		&Eight-Emotion\tnote{1} 
		&Neutral, anger, hate, grief, joy, platonic love,	romantic love, reverence
		&1&L
		&\gls{ecg}, \gls{eda}, \gls{emg}, \gls{resp}  
		\tabularnewline \cline{2-6}
		
		&\multirow{2}{*}{DEAP\tnote{2}}
		&Continuous scale of valence, arousal, liking, dominance, Discrete scale of familiarity
		&32&L
		&\gls{ecg}, \gls{eda}, \gls{eeg}, \gls{emg}, \gls{eog}, \gls{resp}, \gls{skt}, face video (not all subjects)
		\tabularnewline \cline{2-6}
		
		&MAHNOB-HCI\tnote{3}
		&Discrete scale of valence, arousal, dominance, predictability, Emotional keywords
		&27&L
		&\gls{ecg}, \gls{eda} \gls{eeg}, \gls{resp}, \gls{skt}, face and body video, eye gaze tracker, audio
		\tabularnewline\cline{2-6}
		
		&\multirow{2}{*}{StudentLife\tnote{4}}
		&Sleep, activity, sociability, mental well-being, stress, academic performance
		&\multirow{2}{*}{48}&\multirow{2}{*}{F}
		& \gls{acc}, audio, context, GPS, smartphone usage
		\tabularnewline \cline{2-6}
		
		&\multirow{2}{*}{DECAF\tnote{5}}
		&Discrete scale of valence, arousal, dominance
		&	\multirow{2}{*}{30}&\multirow{2}{*}{L} 
		&\gls{ecg}, \gls{emg}, \gls{eog}, MEG, near-infrared face video
		\tabularnewline \cline{2-6}
		
		&\multirow{2}{*}{ASCERTAIN\tnote{6}}
		&Discrete scale of valence, arousal, liking, engagement, familiarity, Big Five 
		& \multirow{2}{*}{58}	&\multirow{2}{*}{L}
		&\gls{ecg}, \gls{eda}, \gls{eeg}, facial activity data (facial landmark trajectories)
		\tabularnewline \hline \hline

		\parbox[t]{2mm}{\multirow{6}{*}{\rotatebox[origin=c]{90}{Stress (S)}}}
		&Driver\tnote{7}	
		& Stress levels: low, medium, high 
		& 24&FC
		& \gls{ecg}, \gls{eda}, \gls{emg}, \gls{resp}
		\tabularnewline \cline{2-6}
		
		&Driver workload\tnote{8}
		& Five different road types, Discrete scale of driver workload	
		& 10		&FC&  \gls{acc}, \gls{ecg}, \gls{eda},  GPS, \gls{skt}, brightness level 
		\tabularnewline\cline{2-6}
		
		&Non-EEG\tnote{9}
		&Four types of stress (physical, emotional, cognitive, none)
		&20	&L
		& \gls{acc}, \gls{eda}, \gls{hr}, \gls{skt}, SpO2
		\tabularnewline\cline{2-6}
		
		& Continuous\tnote{10}
		& Stress levels: none, low, high
		& 21/5		&L/F
		& \gls{acc}, \gls{eda}, \gls{ppg}, \gls{skt}
		\tabularnewline \hline \hline
		
		\parbox[t]{2mm}{\multirow{2}{*}{\rotatebox[origin=c]{90}{E+S}}} 
		&\multirow{2}{*}{WESAD\tnote{11}}
		&\multirow{2}{*}{Three affective states: neutral, fun, stress}
		& \multirow{2}{*}{15}&\multirow{2}{*}{L}
		& \textbf{chest}: \gls{acc}, \gls{ecg}, \gls{eda}, \gls{emg}, \gls{resp}, \gls{skt};
		\textbf{wrist}:  \gls{acc}, \gls{eda}, \gls{ppg}, \gls{skt}
		\tabularnewline
		\hline
	\end{tabular}
	
	\begin{tablenotes}
		\footnotesize{
			References to the datasets:
			\item[1] \citet{2001_Picard},
			\item[2] \citet{2012_Koelstra},
			\item[3] \citet{2012b_Soleymani},
			\item[4] \citet{2014_Wang},
			\item[5] \citet{2015_Abadi},
			\item[6] \citet{2016_Subramanian},
			\item[7] \citet{2005_Healey},
			\item[8] \citet{2013_Schneegass},
			\item[9] \citet{2016_Birjandtalab},
			\item[10] \citet{2016_Gjoreski},
			\item[11] \citet{2018_Schmidt}
		}
	\end{tablenotes}
	\label{tab:avail_data}
	\vspace{-2em}
\end{table}
\textbf{DECAF} (multimodal dataset for DECoding user physiological responses to AFfective multimedia content)\cite{2015_Abadi} is recorded in a laboratory setting with 30 subjects (14 female).
Data recording consisted of two sessions for each subject, presenting music videos and movie clips, respectively.
Considering the music videos, the same set of 40 clips were used as affective stimuli as with the DEAP dataset.
For the movie session, 36 video clips were used as stimuli, from which nine videos can be mapped to each quadrant of the valence-arousal space.
These 36 clips were selected from a larger pool during a pre-study with 42 participants rating videos in the valence-arousal space.
For a detailed description of the final clips, we refer the reader to \citet{2015_Abadi}.
DECAF provides image (near-infrared face videos) and peripheral sensor data (\gls{ecg}, \gls{eog}, \gls{emg}, and MEG (magnetoencephalogram)) from each subject.
A clear limitation of the DECAF dataset is that, due to the MEG recordings, subjects were very restricted in their ability to move.
Therefore, the dataset provides data with a lower noise level than to be expected in real-life settings.

\textbf{ASCERTAIN} (multimodal databASe for impliCit pERsonaliTy and Affect recognitIoN using commercial physiological sensors)\cite{2016_Subramanian} relies on the same 36 video clips for affective stimuli as the DECAF dataset.
ASCERTAIN provides data from 58 subjects (21 female), and includes physiological modalities (\gls{ecg}, \gls{eda}, \gls{eeg}) as well as facial activity data (recorded from a facial feature tracker).
In addition, self-reports including arousal, valence, engagement, liking, and familiarity obtained for each video are included. Moreover, the dataset provides the Big Five personality traits for each subject.
Hence, based on the recorded data, not only models predicting emotions can be created, but also personality traits can be assessed.

\textbf{Driver stress}\cite{2005_Healey} includes data from 24 study participants and the following sensor modalities: \gls{ecg}, \gls{eda}, \gls{emg}, and \gls{resp}.
The dataset was recorded during one \textit{rest} condition and two driving tasks (\textit{highway} and \textit{city}). 
During the two driving tasks the subjects drove for 50 to 90 minutes (depending on the amount of traffic) through city streets and on a highway near Boston, Massachusetts.
Using questionnaires and a score derived from observable events, the three study conditions (\textit{rest, highway, city}) were mapped onto the stress levels low, medium, and high.
Therefore, the dataset facilitates the development of real-life stress monitoring approaches by providing physiological data in relevant scenarios.
However, one important limitation of the dataset is that all sensor data was acquired at low sampling rates (e.g. EMG was sampled at 15.5 Hz).

\textbf{Driver workload} is a dataset with 10 study participants (3 female) by \citet{2013_Schneegass}.
The dataset was recorded while subjects were driving on a defined route of 23.6 km near Stuttgart (Germany), consisting of five different road types (30 km/h zone, 50 km/h zone, highway, freeway, and tunnel).
Furthermore, the authors defined different points of interest along the route, such as freeway exits or roundabouts.
The dataset includes physiological data from the subjects (\gls{ecg}, \gls{eda}, and \gls{skt}) as well as context data collected from a smartphone (GPS, acceleration, and brightness level).
In addition to the five different road types, labels reflecting the participants' perceived workload (from \textit{no workload} to \textit{maximum workload}) are provided.
Therefore, the dataset can be used either to assess mental workload based on physiological data, or the different driving conditions (freeway, highway, inner city) could be mapped onto the classes \textit{medium stress} and \textit{high stress}, as done by \citet{2005_Healey}.

\textbf{Non-EEG}\cite{2016_Birjandtalab} is a dataset containing physiological data (\gls{eda}, \gls{hr}, \gls{skt}, SpO2 - arterial oxygen level, and \gls{acc}) from 20 subjects (4 female).
The dataset was recorded during three different stress conditions (physical, cognitive, and emotional) and a relaxation task.
Physical stress was evoked by asking the subjects to jog on a treadmill at three miles per hour.
In order to elicit cognitive stress, the subjects had to count backwards from 2485 doing steps of seven.
Lastly, emotional stress was triggered by anticipating and watching a clip from a zombie apocalypse movie.
This dataset is particularly interesting as it contains only data from wearable sensors.
Although data was recorded in a lab environment, the subjects were less motion constrained due to the minimally intrusive nature of the sensors, compared to other available datasets.
However, a major limitation of the Non-EEG dataset is the low sampling rate of the employed devices (1 Hz and 8 Hz).
In addition, as no \gls{ecg} or \gls{ppg} data was recorded, the \gls{hrv} information can not be retrieved, a parameter shown to be relevant for stress recognition by various previous work (e.g. \citet{2010_Kreibig}).

\textbf{Continuous stress}\cite{2016_Gjoreski} provides laboratory and real-life data.
In both settings, data was recorded from the Empatica E4 wrist-worn device, including the following sensor modalities: \gls{eda}, \gls{ppg}, \gls{skt}, and \gls{acc}.
The lab study was conducted with 21 subjects.
A mental arithmetic task, which had to be solved under time and social evaluative pressure, served as stressor.
The field study was conducted with 5 subjects, entirely unconstrained, within the subjects' everyday life.
\gls{ema_pl} and manually logging stressful situations served as basis for creating labels for nearly real-life 950 events.
Compared to all the datasets described above, data in particular recorded during everyday life is of interest, as it facilitates the development of continuous stress monitoring approaches.

\textbf{WESAD} (WEarable Stress and Affect Data set) is, to the best of our knowledge, the only publicly available dataset which contains data of subjects experiencing both an emotional and a stress stimulus.
The data collection was conducted with 15 subjects (3 female) in a laboratory setting.
Each subject experienced three conditions: \textit{baseline} (neutral reading task), \textit{fun} (watching a set of funny video clips), and \textit{stress} (being exposed to the \gls{tsst}).
The dataset features physiological and motion data, recorded from both a wrist- and a chest-worn device.
The following sensor modalities are included: \gls{ecg}, \gls{ppg}, \gls{eda}, \gls{emg}, \gls{resp}, \gls{skt}, and \gls{acc}.
Moreover, the high sampling rate (700 Hz) of the chest-worn device should be emphasised.
Overall, WESAD is a fitting dataset for benchmarking affect recognition algorithms based on physiological data.

From \autoref{tab:avail_data} it becomes clear that most available datasets feature \gls{ecg} and/or \gls{eda} recordings (the only exception is the StudentLife dataset which contains no physiological data at all).
This observation coincides with the findings of \Cref{sub:freq_em_sensors}, which indicate that \gls{ecg} and \gls{eda} are the two most frequently employed modalities in the context of affect and stress recognition. 
From the physiological point of view these two modalities are particularly interesting as changes often indicate high arousal states (see \Cref{sec:physio_affect}).
A further observation is that most datasets in \autoref{tab:avail_data} are recorded either in a laboratory setting or in a specific real-life scenario (driving condition). Therefore, the need for further publicly available datasets still exists, especially focusing on everyday life scenarios and including physiological recordings.

\section{Data processing chain}
\label{sec:feat_proc}
In order to associate raw sensor data with different affective states, the standard data processing chain is employed frequently. 
For this purpose, the raw data is first synchronised, filtered, segmented, etc.
A detailed description of these steps is given in \Cref{sub:pre}. 
Once these preprocessing steps are completed, features are computed, aggregating the information present in each signal segment.
An overview of common features, extracted and applied in the wearable affect and stress recognition literature, is given in \Cref{sub:phys_feat}.
Finally, the last step in the standard data processing chain is classification.
During this step a mapping between the feature space and the desired labels (e.g. emotion classes) is learned.
\Cref{sub:class} details common classification methods, applied evaluation frameworks, and the results achieved in related work. 

%
%
%
%
%
%
%

\subsection{Preprocessing and segmentation}
\label{sub:pre}
When multimodal systems are employed, synchronisation of the different raw data streams might be necessary as a first step.
Clear events, e.g. pressing an event marker button or creating double taps, can help to speed up the synchronisation process.
Depending on the transmission protocol of the recorded data, wireless data loss might be an issue.
Different methods exist for handling missing values, as proposed by \citet{2007_Saar}.
Linear interpolation is arguably the simplest of these methods, proven sufficient in most practical settings.
Filtering raw sensor signals, thus removing noise, is another important preprocessing task.
The type of filtering strongly depends on the respective sensor modality.
Therefore, in the following we give an overview of the different filtering and further preprocessing techniques, applied to each modality within the scope of this paper (see \Cref{sub:sensors}).

\begin{enumerate}
	\item \textbf{\gls{acc} Preprocessing:} High frequency artefacts can be removed using a low-pass filter.
	For instance, \citet{2017_Mozos} and \citet{2017a_Gjoreski} considered movement information derived from the \gls{acc} signal in their stress detection systems.
	
	\item \textbf{\gls{ecg} Preprocessing:} In the raw \gls{ecg} signal the R-peaks need to be identified.
	For instance, the Pan and Tompkin's algorithm\cite{1985_Pan} can be applied.
	Once the R-peaks have been detected, the next step is to determine the RR intervals.
	For example, \citet{2015_Hovsepian} present an algorithm to assess the validity of a candidate RR interval.
	Features can then be extracted based on either the identified R-peaks or the RR intervals, as detailed in \Cref{sub:phys_feat}

	\item \textbf{\gls{ppg} Preprocessing:} \citet{2012_Elgendi} gives a detailed description on \gls{ppg} signal preprocessing.
	\gls{ppg} signals are often prone to low-frequency motion artefacts, which can be removed using a high-pass filter.
	For the determination of RR intervals from identified R-peaks, similar algorithms as mentioned with \gls{ecg} preprocessing can be applied.
	
	\item \textbf{\gls{eda} Preprocessing:} Physiological plausible changes in the \gls{eda} signal are commonly in the low-frequency domain.
	Hence, low-pass filtering can be applied to remove high-frequency signal noise.
	After noise removal, the filtered \gls{eda} signal can be detrended by subtracting the low-frequency drift computed by smoothing the signal over a given interval\cite{2012b_Soleymani}.
	As detailed in \Cref{sec:physio_affect}, the \gls{eda} signal consists of two components:
	A slowly varying baseline conductivity referred to as skin conductance level (\gls{scl}) and a higher frequent component called skin conductance response (\gls{scr}).
	\citet{2012_Choi} present an approach to separate the \gls{scl} and \gls{scr} components.

	\item \textbf{\gls{emg} Preprocessing:} Raw \gls{emg} data is often filtered to remove noise.
	For example, \citet{2010_Wijsman} report on a two step procedure.
	First, a bandpass filter, allowing frequencies from 20 to 450 Hz, was applied to the raw signal.
	Then, in order to remove residual power line interference from the bandpass-filtered signal, notch filters were applied.
	The notch filters attenuated the 50, 100, 150, 200, 250 and 350 Hz components of the signal.
	A further common issue with \gls{emg} signals is their contamination by cardiac artefacts.
	\citet{2012_Willigenburg} propose and compare different filtering procedures to remove \gls{ecg} interference from the \gls{emg} signal.
	
	\item \textbf{\gls{resp} Preprocessing:} Depending on the signal quality, noise removal filtering techniques have to be applied. 
	In addition, the raw \gls{resp} signal can be detrended by subtracting a moving average\cite{2009_Khalili}. 
	
\end{enumerate}

Following the preprocessing step in a standard data processing chain, the signal is segmented with a sliding window of (usually) fixed size.
The choice of an appropriate window size is crucial and depends on several aspects, such as the classification task or the applied sensor modalities.
Considering motion signals, relevant patterns usually occur on short time scales.
Therefore, window sizes of $\leq5$ seconds are common, as proved to be useful in the established research field of human activity recognition\cite{2005_Huynh, 2012_Reiss, 2010_Healey}.
The time scale on which physiological responses to an emotional stimulus occur are hard to define.
Hence, considering physiological signals, finding an appropriate window size is a difficult task\cite{2010_Healey}.
Moreover, due to inter-subject and inter-modality (e.g. \gls{emg} vs. \gls{eda}) differences, deciding on an appropriate window size becomes even more challenging.
However, a meta analysis conducted by \citet{2010_Kreibig} found that physiological features are commonly aggregated over fixed window lengths of 30 to 60 seconds.

\subsection{Feature extraction}
\label{sub:phys_feat}

Descriptive features aggregate the information present in signal segments, and serve as input for the classification step of the data processing chain.
Extracted features can be grouped in various ways, such as time- or frequency-domain features, linear or non-linear features, unimodal or multimodal features, etc.
Considering computational complexity, extracted features range from simple statistical features (e.g. mean, standard deviation) to often modality-dependent complex features.
\autoref{tab:feature_list} gives an overview of features commonly extracted and applied in the wearable affect and stress recognition literature.
In the remaining of this section, we give a brief description of feature extraction applied for the different sensor modalities.

\begin{table}[t]
	\centering
	\footnotesize
	\caption{Overview of features commonly extracted and applied in the wearable affect recognition literature.}
	\vspace{-1.2em}
	\begin{tabular}
		{|>{\raggedleft}p{1.cm}
			|>{\raggedright}p{12.0cm}|}
		\hline
		Modality & Features \tabularnewline\hline
		\textbf{\gls{acc}} & \textbf{Time-domain} statistical features (e.g. mean, median, standard deviation, absolute integral, correlation between axes), first and second derivative of acceleration energy\tabularnewline
		&\textbf{Frequency-domain:} power ratio (0-2.75 Hz and 0-5 Hz band), peak frequency, entropy of the normalised \gls{psd}\tabularnewline
		\hline
		&\textbf{Time-domain:} statistical features (e.g. mean, median, 20th and 80th percentile), HR, HRV, statistical features on HRV, number and percentage of successive RR intervals differing by more than 20 ms (NN20, pNN20) or 50 ms (NN50, pNN50), pNN50/pNN20 ratio\tabularnewline
		& \textbf{Frequency-domain:} ultra low (ULF, $0-0.003\,Hz$), very low (VLF, $0.003-0.03\, Hz$), low (LF, $0.03-0.15\, Hz$), and high (HF, $0.15-0.4\,Hz$) frequency bands of HRV, normalised LF and HF, LF/HF ratio \tabularnewline
		\textbf{\gls{ecg}/ \gls{ppg}}&\textbf{Non-linear:} Lyapunov exponent, standard deviations ($SD_{1}$ and $SD_{2}$) from Poincar\'e plot,  $SD_{1}/SD_{2}$ ratio, sample entropy\tabularnewline
		&\textbf{Geometrical:} triangular interpolation of R peak intervals, histogram (TINN)\tabularnewline
		&\textbf{Multimodal:} respiratory sinus arrhythmia (\gls{rsa}), respiration-based HRV decomposition\tabularnewline
		\hline
		\textbf{\gls{eda}}& \textbf{Time-domain:} statistical features (mean, standard deviation, min, max, slope, average rising time, mean of derivative, etc.)\tabularnewline
		&\textbf{Frequency-domain}: 10 spectral power in the 0-2.4 Hz bands\tabularnewline
		&\textbf{\gls{scl} features}: statistical features, degree of linearity\tabularnewline
		&\textbf{\gls{scr} features}: number of identified SCR segments, sum of SCR startle magnitudes and response durations, area under the identified SCRs\tabularnewline
		\hline
		
		\textbf{\gls{emg}}&\textbf{Time-domain:} statistical features, number of myoresponses\tabularnewline 
		&\textbf{Frequency-domain:} mean and median frequency, energy\tabularnewline 
		\hline
		
		\textbf{\gls{resp}} & \textbf{Time-domain:} statistical features (e.g. mean, median, 80th percentile) applied to: breathing rate, inhalation (I) and exhalation (E) duration, ratio between I/E, stretch, volume of air inhaled/exhaled\tabularnewline
		&\textbf{Frequency-domain:} mean power values of four subbands (0-0.1 Hz, 0.1-0.2 Hz, 0.2-0.3 Hz and 0.3-0.4 Hz)\tabularnewline
		&\textbf{Multimodal:} respiratory sinus arrhythmia (\gls{rsa})\tabularnewline
		\hline
		
		\textbf{\gls{skt}}& \textbf{Time-domain:} statistical features (e.g. mean, slope), intersection of the y-axis with a linear regression applied to the signal\tabularnewline
		\hline
	\end{tabular}
	\label{tab:feature_list}
	\vspace{-1.8em}
\end{table}

From the domain of human activity recognition, a large set of features based on acceleration data is known.
These features are often also employed in the context of affect and stress recognition.
Statistical features (mean, median, standard deviation, etc.) are often computed for each channel ($x, \, y, \, z$) separately and combined.
\citet{2007_Prakka} showed that the absolute integral of acceleration can be used to estimate the metabolic equivalent of physical activities, which can be an interesting feature for affect recognition as well.
Moreover, \citet{2017_Mozos} used the first and second derivative of the accelerometer's energy as feature, e.g. to indicate the direction of change in activity level.
Considering frequency-domain features, the power ratio of certain defined frequency bands, the peak frequency, or the entropy of the Power Spectral Density (PSD) have been applied successfully. 

Based on \gls{ecg} and \gls{ppg} data, various features indicating physiological changes in the cardiac cycle can be computed.
First, commonly the \gls{hr} (beats per minute) is derived.
Based on the location of the R-peaks (or the systolic peak in the \gls{ppg} signal) the \gls{ibi} can be computed. 
The \gls{ibi} serves as a new time series signal, from which various \gls{hrv} features can be derived, both in time- and frequency-domain.
For instance, from the \gls{ibi} the number and percentage of successive RR intervals differing by more than a certain amount of time (e.g. 20 or 50 milliseconds) can be computed. 
These feature are referred to as NNX and pNNX, where X is the time difference threshold in milliseconds.
Based on the Fourier-transformation of the \gls{ibi} time series, various frequency-domain features can be computed, wich reflect the sympathetic and parasympathetic activities of the autonomous nervous system (\gls{ans}).
Four different frequency bands are established in this respect\cite{2016_Rubin}.
The ultra low frequency (ULF) and very low frequency (VLF) bands range from 0 to 0.003 Hz and from 0.003 to 0.03 Hz, respectively.
The low frequency (LF) band, ranged between 0.03 and 0.15 Hz, is believed to reflect mostly the sympathetic activity (with some parasympathetic influence) of the \gls{ans}.
In contrast, the high frequency (HF) band, ranged from 0.15 to 0.4 Hz, is associated with parasympathetic activities\cite{2016_Rubin}.
Therefore, the LF/HF ratio might be a representative feature of the sympathetic to parasympathetic influence on cardiac activity, thus might be a good stress indicator\cite{2005_Healey}.
Non-linear features have been derived from the ECG signal as well: maximal Lyapunov exponent, standard deviations ($SD_{1}$ and $SD_{2}$) along major axes of a Poincar\'e plot, the $SD_{1}/SD_{2}$ ratio, sample entropy etc.
For a detailed description of these features, see \citet{2016_Rubin}.
\citet{2012_Valenza}, with the goal to detect five levels of valence and arousal, compared the performance of a quadratic discriminant classifier which based its decision on a set of linear and non-linear features. 
Their results indicate that non-linear features are able to improve classification results significantly.
Another class of features based on the cardiac cycle are referred to as geometrical features.
An example is the triangular interpolation index (TINN)\cite{1996_Malik, 2012_Valenza, 2016_Rubin}: a histogram of the RR intervals is computed and a triangular interpolation performed.
Finally, respiration is known to have an impact on the \gls{ecg} signal.
The actual cause (linked to chest expansion/contraction or driven by a brainstem circuit) of the observed effect is still under debate.
However, there exist different methods for quantifying the effect of the respiration pattern on the variability in the RR intervals.
For instance, the \gls{rsa} can be calculated, which is a combined feature of \gls{resp} and \gls{ecg}\cite{2015_Hovsepian}.
In addition, \citet{2012_Choi} propose a method of decomposing the \gls{hrv} into a respiration- and a stress-driven component.
Overall, for a more detailed description of features based on the cardiac cycle, we refer to \citet{1996_Malik}.

Considering the EDA signal, basic statistical features (e.g. mean, standard deviation, min, max) are commonly used\cite{2010_Setz}.
\citet{2012_Koelstra} provides a list of further EDA-related statistical features, such as average rising time or the average decrease rate during decay time.
\citet{2012_Koelstra} also extracted frequency-domain features from the EDA signal: 10 spectral power values in the 0-2.4 Hz frequency bands.
After separating the EDA signal into \gls{scl} and \gls{scr}, further features are extracted from each component.
Since the \gls{scl} component represents a slowly varying baseline conductivity, its degree of linearity proved to be a useful feature\cite{2012_Choi}.
Considering the \gls{scr} component, the identified SCR segments are counted and further statistical features derived: sum of the \gls{scr} startle magnitudes and response durations, area under the identified \gls{scr}s\cite{2005_Healey}, etc.
The \gls{scr}-related features were found to be particularly interesting as they are closely linked to high arousal states\cite{2008_AKim}.

From the \gls{emg} signal, various time- and frequency-domain features can be extracted.
\citet{2012_Christy} computed statistical features such as mean, median, standard deviation, and interquartile ranges on the \gls{emg} data.
The median \gls{emg} feature of the trapezius muscle was the highest ranked feature of a binary valence classifier.
Other researchers used frequency-based features such as peak \cite{2016_Kollia} or mean frequencies\cite{2013_Wijsman}.
Another frequently used feature is the signal energy of either the complete signal\cite{2012_Koelstra} or specific frequency ranges (e.g. 55-95 Hz, 105-145 Hz)\cite{2015_Abadi}.
\citet{2013_Wijsman} performed a reference voluntary contraction measurement to compute a personalised \gls{emg} gap feature.
This feature is defined as the relative time the \gls{emg} amplitude is below a specific percentage of the amplitude of the reference measurements.

\citet{2012b_Soleymani} pointed out that slow respiration is linked to relaxation. 
In contrast, irregular and quickly varying breathing patterns correspond to more aroused states like, anger or fear\cite{2008_AKim, 2006_Rainville}.
Therefore, different respiration patterns can provide valuable information for the detection of affective states.
\citet{2011_Plarre} describe a number of time-domain features which aggregate information about breathing cycles: breathing rate, inhalation (I) and exhalation (E) duration, ratio between I/E, stretch (the difference between the peak and the minimum amplitude of a respiration cycle), and the volume of air inhaled/exhaled.
Considering frequency-domain features, \citet{2014_Kukolja} used mean power values of four frequency subbands (0-0.1 Hz, 0.1-0.2 Hz, 0.2-0.3 Hz and 0.3-0.4 Hz) in order to classify different types of emotions.
Moreover, as a multimodal feature (derived from both \gls{ecg} and respiration), \gls{rsa} is commonly used\cite{2011_Plarre, 2015_Hovsepian}.

Changes in body temperature might be attributed to the 'fight-or-flight' response (see \Cref{sec:physio_affect}).
During this physiological state, the blood flow to the extremities is restricted in favour of an increased blood flow to vital organs.
Hence, temperature-based features can be relevant indicators for e.g. a severe stress response.
\citet{2017a_Gjoreski}, for instance, extract the mean temperature, the slope, and the intersection of a linear regression line with the y-axis as features. 

\subsection{Classification}
\label{sub:class}

Classification of an affective state is either performed using statistical (e.g. ANOVA) or machine learning (e.g. SVM, kNN) techniques.
For both types of analysis, a subset of the features presented in \Cref{sub:phys_feat} can be used as input.
Since statistical analysis plays only a minor role in the investigated wearable affect and stress recognition literature, we focus in this section on classification utilising machine learning techniques.
\autoref{tab:lit_meth} compares the studies presented in \autoref{tab:aim}, using the same chronological order.
This comparison focuses in particular on the employed classification algorithms and evaluation methods, as well as the achieved classification performance.
The classification performance is, if possible, reported as accuracy, indicating the overall percentage of correctly classified instances.
Furthermore, \autoref{tab:lit_meth} details the number of defined affective classes, the location of each study, and the number of study participants.
The rest of this section discusses each of these aspects in detail.
\glsreset{svm}
\glsreset{lda}
\glsreset{rf}
\glsreset{knn}
\glsreset{ar}


The algorithm column in \autoref{tab:lit_meth} indicates that the \gls{svm} is the most frequently employed classification algorithm (applied in 43\% of the studies).
This is to some degree surprising as the \gls{svm} requires careful adjustment of the kernel size  $\gamma$ and the trade-off parameter $C$. 
For this adjustment the recorded data has to be split into a training, validation, and test set.
The best set of hyperparameters can be found by performing a grid-search\cite{2015_Hovsepian,2017_Mozos}, evaluating the current hyperparameter on the validation set.
The performance of the final model is then evaluated on the test set.
Hence, when applying a \gls{svm}, it is important to report the final test error (and \textit{not} the validation error).
The second and third most popular classifiers are \gls{knn} and \gls{lda}, being applied in 27\% and 14\% of the studies, respectively.
\gls{lda} and \gls{knn} require only little hyperparameter tuning.
Hence, both can be applied (almost) in an off-the-shelf way.
Concluding from \autoref{tab:lit_meth}, ensemble methods (e.g. boosting or random forest) were employed less frequently.
This is contradictory to the fact that ensemble methods have proved to be strong classifiers.
\citet{2014_Delgado} evaluated 179 classifiers on more than hundred different datasets and found that the \gls{rf} family 'is clearly the best family of classifiers'\cite{2014_Delgado}.
In the affect recognition community, \citet{2016_Rubin} for instance employed the \gls{rf} classifier, reaching 97\% and 91\% accuracy on classifying \textit{panic} and \textit{pre-panic} states, respectively.
In addition, boosting was found to be a strong classifier\cite{2014_Delgado}, and Leo Breiman even considered it to be the 'best off-the-shelf classifier in the world'\cite{2000_Friedman}.
\citet{2017_Mozos} applied the AdaBoost method to detect stress, reaching an accuracy of 94\%.
\citet{2014_Delgado} also found \gls{nn} to be among the top-20 classifiers.
\citet{2004_Haag} and \citet{2016_Jaques} used \gls{nn}, in the form of multi-layered perceptrons, to detect different affective states.
Convolutional neural networks (CNN) or \gls{lstm} based classification techniques, which are becoming popular in the field of human activity recognition\cite{2016_Hammerla, 2017_Munzner}, have not found broad application in the domain of wearable affect and stress recognition yet.
\citet{2013_Martinez} compare the performance of learned and hand-crafted feature sets to detect the affective states \textit{relaxation, anxiety, excitement} and \textit{fun}.
The learned features were extracted using a set of convolutional layers, and the final classification step was performed using a single-layer perceptron.
The experiments of \citet{2013_Martinez} indicate that learned features are able to improve classification performance over results which rely on manually constructed statistical feature extraction.

Considering the defined affective classes, binary classification tasks are common according to \autoref{tab:lit_meth}.
Even if the study protocol aimed at eliciting different emotions, the classification problem was formulated in a binary fashion.
A frequent task is to separate high and low valence/arousal based on physiological data\cite{2012_Agrafioti,2015_Abadi}.

\glsreset{cv}
\glsreset{loo}
\glsreset{loso}
\glsreset{loto}
\begin{threeparttable}[t!]
	\footnotesize{
		\centering
		\caption{Comprehensive comparison of the analysed wearable affect and stress recognition literature.\newline
			\footnotesize{If not stated differently, scores are reported as (mean) accuracy. Abbreviations: Location (Loc), Lab (L), Field (F), Field with constraint (FC),
				\gls{cv}, \gls{loo}, \gls{loso}, \gls{loto}, Arousal (AR), Valence (VA), Dominance (DO), Liking (LI),
				AdaBoost (AB), Bayesian Network (BN), Gradient Boosting (GB), Linear Discriminant Function (LDF), Logistic Regression (LR), Naive Bayes (NB), Neural Net (NN), Passive Aggressive classifier (PA), Random Forest (\gls{rf}), Decision/Regression/Function Tree (DT/RT/FT), Ridge Regression (RR)}\vspace{-1em}}
		\begin{tabular}
			{|>{\raggedright}p{2.2cm}
				|>{\raggedright}p{2.5cm}
				|>{\centering}p{0.9cm}
				|>{\centering}p{.8cm}
				|>{\centering}p{.7cm}
				|>{\raggedright}p{1.3cm}
				|>{\raggedright}p{3cm}|}
			\hline
			Author & Algorithm &  Classes & Loc. &	Sub. & Validation & Accuracy\tabularnewline
			\hline
			\citeauthor{2001_Picard}
			&kNN &8&L& 1&\acrshort{loo}&81\%
			\tabularnewline \hline 
			
			\citeauthor{2004_Haag}
			&NN&contin.&L&1&3-fold split& AR: <96\%, VA: <90\%
			\tabularnewline\hline	
			
			\citeauthor{2004_Lisetti}
			&kNN; LDA; NN&6&L&14&\acrshort{loo}&72\%; 75\%; 84\%
			\tabularnewline \hline
			
			\citeauthor{2005_Liu}
			&kNN; RT; BN; SVM &5&L&15&\acrshort{loo}& 75\%; 84\%; 74\%; 85\%
			\tabularnewline \hline
			
			\citeauthor{2005_Wagner}
			&kNN; LDF; NN & 4 &L&1&\acrshort{loo}&81\%; 80\%; 81\%
			\tabularnewline \hline
			
			\citeauthor{2005_Healey}
			&LDF&3&FC&24&\acrshort{loo}&97\%
			\tabularnewline \hline
			
			\citeauthor{2007_Leon}
			&NN&3&L&8+1&\acrshort{loso}&71\%
			\tabularnewline\hline
			
			\citeauthor{2008_Zhai}
			&NB; DT; SVM&Binary&L&32&20-fold \acrshort{cv}&79\%; 88\%; 90\%
			\tabularnewline \hline

			\citeauthor{2008_Kim}
			&LR&Binary&FC&53&5-fold \acrshort{cv}& $\sim\,63\%$
			\tabularnewline\hline
			
			\multirow{2}{*}{\citeauthor{2008_AKim}}
			&LDA&4&L&3&\acrshort{loo}& subject dependent: 95\%,\\subject independent: 70\%
			\tabularnewline\hline
			
			\citeauthor{2008_Katsis}
			&SVM; ANFIS&4&L&10&10-fold \acrshort{cv}& 79\%;77\%
			\tabularnewline\hline
			
			\multirow{ 2}{*}{\citeauthor{2009_Calvo}}
			&FT; NB; BN; NN; LR, SVM&8&L&3&10-fold \acrshort{cv}&one subject: 37\%-98\%, \\all subjects: 23\%-71\%
			\tabularnewline\hline
			
			\citeauthor{2009_Channel}
			&LDA; QDA; SVM&3&L&10&\acrshort{loso}& <50\%; <47\%; <50\%,\\ Binary: <70\%
			\tabularnewline\hline 
			
			\citeauthor*{2009_Khalili}
			&QDA&3&L&5&\acrshort{loo}&66.66\%
			\tabularnewline\hline

			\citeauthor{2010_Healey}
			&BN; NB; AB; DT&Binary&F&19&10-fold \acrshort{cv}& None\tnote{{2}}
			\tabularnewline\hline
			
			\citeauthor{2011_Plarre}					
			&DT; AB; SVM/HMM&Binary&L/F&21/17& 10-fold \acrshort{cv}& 82\%; 88\%; 88\%/ 0.71\tnote{3}
			\tabularnewline\hline
			
			\citeauthor{2011_Hernandez}
			&SVM&Binary&F&9&\acrshort{loso}&73\% 
			\tabularnewline\hline
			
			\citeauthor{2012_Valenza}
			&QDA&5&L&35&40-fold CV& >90\%
			\tabularnewline \hline
			
			\citeauthor{2012_Hamdi}
			&ANOVA&6&L&16&-&None \tnote{4}
			\tabularnewline\hline
			
			\citeauthor{2012_Agrafioti}
			&LDA&Binary&L&31&\acrshort{loo}&Active/Pas AR: 78/52\% \\Positive/Neg VA: <62\%
			\tabularnewline\hline
			
			\citeauthor{2012_Koelstra}
			&NB&Binary&L&32&\acrshort{loso}&AR/VA/LI: 57\%/63\%/59\%
			\tabularnewline \hline
			
			\citeauthor{2012b_Soleymani}
			&SVM&3&L&27&\acrshort{loso}&VA: 46\%, AR: 46\%
			\tabularnewline\hline
			
			\citeauthor{2013_Sano}
			&SVM, kNN&Binary&F&18&10-fold \acrshort{cv}&<88\%
			\tabularnewline\hline
			
			\citeauthor{2013_Martinez}
			&NN	& 4\tnote{1} &L&36&3-fold \acrshort{cv}& learned features: <75\%,\\hand-crafted: <69\%
			\tabularnewline \hline
			
			\citeauthor{2014_Valenza}
			&SVM&Binary&L&30&\acrshort{loo}& VA: 79\%, AR: 84\%
			\tabularnewline\hline
			
			\citeauthor{2014_Adams}
			&GMM&Binary&F&7&-&74\%
			\tabularnewline\hline
			
			\citeauthor{2015_Hovsepian}
			&SVM/BN&Binary&L/F&26/20&\acrshort{loso}& 92\%/>40\%
			\tabularnewline\hline
			
			\citeauthor{2015_Abadi}
			&NB, SVM&Binary&L&30&\acrshort{loto}&VA/AR/DO: 50-60\%
			\tabularnewline \hline
			
			\citeauthor{2016_Rubin}
			&PA; GB; DT; RR; SVM; RF; kNN; LR &Binary&F&10&10-fold \acrshort{cv}&Bin. panic: 73\% - 97\% \\ Bin. pre-panic: 71\% - 91\%
			\tabularnewline \hline
			
			\citeauthor{2016_Jaques}
			&SVM; LR; NN; &Binary&F&30&5-fold \acrshort{cv}&<76\%; <86\%; <88\%
			\tabularnewline\hline
			
			\citeauthor{2016_Rathod}
			&Rule based&6&L&6&-&<87\%\tabularnewline \hline
			
			\citeauthor{2016_Zenonos}
			&kNN; DT; RF&5&F&4&\acrshort{loso}&58\%; 57\%; 62\%
			\tabularnewline\hline
			
			\citeauthor{2016_Zhu}
			&RR&1&F&18&\acrshort{loso}&$0.24\pi \approx 43^\circ$\tnote{5}
			\tabularnewline\hline 
			
			\citeauthor{2016_Birjandtalab}
			&GMM&4&L&20&-& <85\%\tabularnewline \hline
			
			\citeauthor{2017a_Gjoreski}
			& SVM; RF; AB; kNN; BN; DT &3/Binary&L/F&21/5&\acrshort{loso}&<73\%/ <90\%
			\tabularnewline\hline
			
			\citeauthor{2017_Mozos}
			&AB; SVM; kNN &Bin.&L&18&\acrshort{cv}&94\%; 93\%; 87\%
			\tabularnewline\hline		
			
			\citeauthor{2018_Schmidt}
			& DT; RF kNN; LDA; AB &3/Binary&L&15&\acrshort{loso}& <80\%/<93\%
			\tabularnewline\hline
		\end{tabular}
		\label{tab:lit_meth}
		\begin{tablenotes}
			\footnotesize{
				\item[1] Given as pairwise preferences,
				\item[2] DT overfit, other classifiers performed worse than random guessing,
				\item[3] Correlation between self-reported and output of model,
				\item[4] No significant	differences could be found between the affective states,
				\item[5] Mean absolute error of mood angle in circumplex model
				\vspace{1em}
			}
		\end{tablenotes}
	}
\end{threeparttable}

Considering location, three different types of studies are distinguished: \textit{lab} (L), \textit{field} (F), and \textit{field with constraint} (FC) studies.
Studies conducted in a vehicle on public roads are referred to as FC studies, as subjects are constrained in their movement.
Most, 24 out of 37, studies presented in \autoref{tab:lit_meth}, solely base their results on data recorded in a lab setting.
The popularity of lab studies is easily explained, since they allow to design a protocol that elicits a set of target emotions and the same protocol can be applied to multiple subjects.
However, models trained on data gathered in a constrained environment are prone to errors in a more generalised setting.
Therefore, and due to recent advances in mobile and sensor technology, field studies have become more frequent over the past years.
In order to reliably predict the affective state of a user in everyday life, this 'out of the lab and into the fray'\cite{2010_Healey} trend is certainly desirable.
For example, several recent work aim to detect stress in lab and real life scenarios\cite{2017a_Gjoreski, 2015_Hovsepian, 2011_Plarre}.
The authors of these work conducted lab and field studies, and evaluated their algorithms trained on lab data in unconstrained settings.
The results indicate that stress detection in completely unconstrained environments is a feasible task.
Finally, considering the number of study participants there is a large variation, ranging between a single subject and up to 68 subjects.
Clearly, a large and diversified subject pool is desirable, as it would allow to develop generalised models for affect recognition.

\glsreset{cv}
\glsreset{loo}
\glsreset{loso}
\glsreset{loto}
\autoref{tab:lit_meth} indicates that n-fold ($n \in [3, 5, 10, 20,40 ]$) \gls{cv} is frequently employed as validation method.
Following this method, the dataset is randomly partitioned into $n$ equal size subsets.
Then, $n-1$ subsets are used as training data and the remaining subset as test data.
This procedure is repeated $n$ times, so that each of the $n$ subsets is used exactly once as test data.
In case the trained model requires hyperparameter tuning, part of the training data can serve as validation set in each iteration.
\gls{loo} \gls{cv} is also used in several of the listed studies in \autoref{tab:lit_meth}.
This is a specific version of the n-fold \gls{cv} procedure, where $n$ equals the number of available samples.
However, general cross-validation techniques lead to subject-dependent evaluation results.
In order to simulate subject independency, and thus to obtain more realistic results for real-life deployment, the validation method \gls{loso} \gls{cv} should be applied.
For this purpose, the algorithm under consideration is trained on the data of all but one subject.
The data of the left-out subject is then used to evaluate the trained model.
Repeating this procedure for all subjects in the dataset gives a realistic estimate of the model's generalisation properties.
A slightly different type of validation was performed by \citet{2015_Abadi}: \gls{loto} \gls{cv}.
During \gls{loto} CV, the model is trained on the data of all subjects but leaving one trial/stimulus (e.g. video) aside.
The trained algorithm is then evaluated on the left-out data, and the procedure is repeated for each trial.
Overall, the \gls{loso} procedure is nowadays widely accepted and applied, as indicated by several examples listed in \autoref{tab:lit_meth}.
From the results shown here, it can be concluded that using the \gls{loso} validation method leads to lower classification scores than applying n-fold or \gls{loo} \gls{cv}.
However, only \gls{loso} provides the information on how good the trained model is able to perform on completely unseen data (e.g. data of a new user), and hence this method should be used.

\glsreset{sns}
The affect and stress recognition approaches presented in \autoref{tab:lit_meth} report accuracies between 40\% and 95\%.
Due to the lack of benchmarking datasets, the results obtained in different studies are hard to compare.
In general, performance results obtained in lab studies are on average higher than the ones obtained in field studies.
\citet{2015_Hovsepian}, who conducted both a lab and a field study, report on a 92 \% mean accuracy in detecting stress based on lab data. 
However, when field data is considered, the accuracy drops to $62\, \%$.
Moreover, \citet{2010_Healey} conducted a field study and trained different classifiers on the collected data, but none of them was able to perform better than random guessing.
This indicates that training classifiers detecting different affective states based on field data is difficult.
From \citet{2015_Dmello} it is known that multimodal classification methods reach on average higher performance than systems relying only on unimodal input.
This coincides with \autoref{tab:aim}, from which also the conclusion can be drawn that almost all investigated studies utilised multimodal sensory setups.
Considering the accuracy of classifiers detecting high/low arousal and high/low valence separately it becomes apparent, that on average arousal is classified more reliably\cite{2004_Haag,2012_Valenza,2012_Agrafioti,2015_Abadi}.
High arousal states are from a physiological point of view directed by the \gls{sns} (see \Cref{sec:physio_affect}).
As the physiological changes directed by the \gls{sns} are quite distinct (e.g. increase in \glsdesc{hr}, sweating), detecting high arousal states using physiological indicators is a feasible task.
In contrast, detecting changes in a subject's valence based on physiological data is a more challenging task.

The performance of standard machine learning classifiers depend strongly on the input features.
Hence, the benefits of a careful feature selection can be threefold.
First, feature selection can improve classification results.
Second, feature selection can help to identify cost-effective predictors.
Third, it provides a better understanding of the processes generating the data\cite{2003_Guyon}.
According to \citet{2003_Guyon}, feature selection methods are grouped into filter-based methods, wrappers, and embedded methods.
Filter-based methods select a subset of features based on statistical criteria and do not take the used classifier into account. 
In contrast, wrappers (e.g. sequential feature selection) treat the learning algorithm as black box and assess the quality of a subset of features based on the classification score.
Finally, embedded methods perform variable selection during training.
Hence, the selection is commonly specific to the used classifier\cite{2003_Guyon}.
Feature selection methods also find application in affective computing.
\citet{2008_AKim}, for instance, perform feature selection to improve the classification. 
\citet{2012_Valenza} used \gls{pca} to project the features onto a lower dimensional space. 
This linear method has the advantage that the features are condensed with only a minimal loss of information.
For a detailed review of feature selection methods see \citet{2003_Guyon}.


\section{Discussion and outlook}
\label{sec:discussion}

Based on the previous sections, we would like to go one step further and identify key challenges and opportunities in wearable affect and stress recognition.
We will focus on the following issues in this section:
a) valence detection,
b) hardware,
c) datasets,
d) algorithmic challenges,
and e) long-term reasoning.

\textbf{Valence detection}:
From \Cref{se:stress_emo} and \Cref{sec:physio_affect}, the link between physiological changes and the arousal axis of the circumplex model became apparent.
Hence, it is not surprising that approaches of stress detection and arousal assessment in \autoref{tab:lit_meth} reach high accuracy.
However, valence-related changes in human physiology are more subtle and therefore more difficult to detect.
This explains the generally lower accuracy of the valence detection systems in \autoref{tab:lit_meth}.
In some studies\cite{2012_Koelstra,2015_Abadi}, smiles and other facial expressions, which are directly connected to valence, are recorded using facial \gls{emg} or \gls{eog}.
However, this procedure is not applicable in everyday life due to practical considerations.
One possibility to improve the assessment of valence is to incorporate contextual data into the classification process.
This contextual information can range from audio samples (e.g. laughter), information about the sleep cycle and its quality, to calendar meta data or text (e.g. emails/chat).
Following for instance \citet{2015_Sano}, the regularity of sleep and duration has a very strong impact on the mood of a person and is a strong feature to predict the morning mood. 

\textbf{Hardware}:
The setups used to record physiological data in \glsdesc{ar} studies are often either watch-like (e.g. Empatica E4\cite{E4}), chest-belt (e.g. AutoSense\cite{2011_Ertin}), or stationary devices (e.g. BioPac systems\cite{BioPac}).
Recent progress in flexible electronics enabled the development of sensor patches (e.g. Vivalnk\cite{Vivalnk}), which have not been applied in many \glsdesc{ar} studies yet.
Furthermore, sensors and processing units can be integrated into fabric (for a comprehensive summary see \citet{2015_Reiss}).
These technologies offer an increased wearing comfort, potential new measurement positions\cite{2017_Lonini}, and are less intrusive.
Hence, they certainly deserve more attention in wearable \glsdesc{ar}.
In addition to the traditionally employed set of modalities (\gls{ecg}, \gls{eda}, etc.), the merits of other sensors should be investigated. 
First, considering the cardiac system, stress has been related to changes in blood pressure\cite{2000_Vrijkotte}.
Hence, incorporating data representing a blood pressure correlate (e.g. pulse wave transit time\cite{2012_Gesche}) could enable more reliable stress detectors.
Second, body microphones placed on the subject's chest or abdomen could provide further insights into the cardiac\cite{2010_Pandia}, respiration, and digestive system.
Third, the chemical composition of perspiration could provide further information about the physiological state of a person.
Hence, integrating chemical-electrophysiological sensors\cite{2016_Gao,2016_Imani} in \glsdesc{ar} studies has the potential to create new insights into the physiology of affective states.
Finally, as already mentioned above, contextual information about the user could help to improve the classification of the affective state of the person wearing the device.
The sources of contextual data are nearly unlimited and range from ambient audio data to video streams provided by devices like smart glasses (e.g. Google Glass).
These sources could be used to classify the surroundings of a user and the affective state of other nearby persons as well.

\textbf{Datasets}:
The wearable \glsdesc{ar} community lacks publicly available datasets, frequently used for benchmarking. 
In order to generate statistically meaningful results, a representative cohort of subjects is desirable.
However, most \glsdesc{ar} studies target students or research staff, which are likely to represent a homogeneous group\cite{2001_Peterson}.
In order to mitigate this selection bias, studies could recruit subjects from different social groups (gender, age, etc.).
The available datasets (see \Cref{sec:data_set}) already feature multiple modalities.
However, measuring physiological changes in a redundant fashion (e.g. using \gls{ecg} and \gls{ppg}) or using the same modality on various locations (e.g. wrist and torso) would facilitate a direct comparison of the signals.
Studies on wearable emotion detection commonly elicit and detect multiple emotional states\cite{2001_Picard,2012_Koelstra}.
In contrast, stress detection systems mainly target binary problems (\textit{stress} vs. \textit{no-stress}).
In our opinion, robust \glsdesc{ar} systems should be trained on datasets like WESAD\cite{2018_Schmidt} which include stress and other affective states.
Up-to-date \glsdesc{ar} research based on wearables mainly focus on lab studies.
For benchmarking and exploitative studies, lab data is a good starting point.
However, real-life \glsdesc{ar} systems need to be trained on data acquired in an unconstrained environments. 
Hence, we hope that the observed trend towards field studies (see \autoref{tab:lit_meth}) continues.
To support this trend, we provided in \Cref{sub:guide} practical guidelines on ground truth generation in field studies.

\textbf{Algorithmic challenges}:
The way humans perceive and react to an affective stimulus is subject dependent. 
This highlights the importance of personalisation.
However, the current state-of-the-art in wearable \glsdesc{ar} makes little use of personalisation methods.
One way to account for the subjective nature of affective states is to utilise online learning.
Following this idea, a general model could be deployed, which is then customised.
Customisation could happen for instance via an active labelling approach, where the user is occasionally asked to provide labels.
In addition, semi-supervised or even unsupervised training methods could be used. 
To the best of our knowledge, these methods have not found application in wearable \glsdesc{ar} research yet.
\glsreset{nn}
In most studies presented in \autoref{tab:lit_meth} classical machine learning algorithms (e.g. kNN, SVM) were applied.
In \glsdesc{har}\cite{2015_Yang}, audio analysis\cite{2013_Wollmer}, or stock return forecasting\cite{2005_Enke}, which all deal with time series data, (deep) artificial \gls{nn} proved to be powerful classifiers.
Using \gls{nn} makes feature engineering obsolete, as via backpropagation features are learned.
From a methodical point of view \gls{nn} offer interesting approaches to transfer\cite{2016_Roggen} or semi-supervised\cite{2010_Vincent} learning.
Deep neural networks require a large amount of training data and are known to be resource intensive. 
Hence, deployment on an embedded device is an open research question.
Due to strong interest in \gls{nn} from both academia and industry, we are confident that resource-related issues will be solved in near future.

\textbf{Long-term reasoning}:
Image-based \glsdesc{ar} systems can only perform a temporal- and spatial-limited assessment of the user's state (e.g. while driving\cite{affectiva}). 
In contrast, wearable-based \glsdesc{ar} systems detect the user's affective state continuously and ubiquitously. 
This can be used for a deeper analysis, providing reasoning for certain affective states or behavioural patterns.
First approaches of long-term reasoning were presented by \citet{2017a_Gjoreski} and in the HappyMeter App\cite{2017_Budner}. 
The latter investigated correlations between affective states and environmental conditions (e.g. temperature, wind, humidity) or persons nearby.
Visualising this information can increase awareness of specific situations (e.g. showing locations where the user is stressed).
Essential for this correlation analysis, is contextual information.
We see large potential for this research direction, as the reasoning methods presented above are still in an early stage.
\\

The aim of this review was to provide a broad overview and in-depth understanding of the theoretical background, methods, and best practices of wearable affect and stress recognition.
Currently, there is a strong trend to small, lightweight, affordable, and wearable electronic gadgets.
These devices can be used for sensing, storing and data processing\cite{2012_Miller} and hence offer an ideal platform for enhanced \glsdesc{ar} systems.
There is a wide range of applications for such systems, in particular in the consumer and healthcare domain.
From a healthcare point of view, wearable \glsdesc{ar} systems could, for instance, help to ubiquitously monitor the state of patients with mental disorders (e.g. depression).
This data could provide valuable insights for therapists, promoting behaviour change interventions\cite{2015_Kanjo}.
Furthermore, these systems could facilitate the development of tele-mental\cite{2014_Chan} and tele-medical applications.
Wearable \glsdesc{ar} systems could improve self monitoring, provide users with a better understanding of their affective states, and support behavioural changes.
Beyond these health-related applications, \glsdesc{ar} systems could be used in urban planning\cite{2012_Bergner} or to improve human-machine interfaces.
Despite the impressive progress made in recent years, the applications mentioned above are still under research and not available for customers.
We are convinced that robust and personalised \glsdesc{ar} systems applicable in everyday life could provide many users with an added value.
Hence, we encourage the community to support and address the remaining challenges.

\bibliographystyle{ACM-Reference-Format-Journals}
\bibliography{the_bib_compr}
\end{document}